\pdfoutput=1
\documentclass{llncs}

\usepackage{amsmath,amssymb}

\usepackage[]{graphics}
\usepackage{tikz}

\usepackage{cuted}
\usepackage{pdflscape}
\usepackage{rotating}
\usepackage{authblk}
\usepackage{appendix}

\usepackage{graphicx}
\usepackage{booktabs} 

\usepackage{mathtools}

\allowdisplaybreaks

\usepackage{textcomp,marvosym}

\usepackage{xargs}                      

\usepackage{array}

\usepackage{listings} 

\usepackage{subcaption}
\usepackage[british]{babel}
\usepackage{hhline}
\usepackage{multirow}

\usepackage{algorithm}

\usepackage[noend]{algpseudocode}
\algdef{SE}[DOWHILE]{Do}{doWhile}{\algorithmicdo}[1]{\algorithmicwhile\ #1}%

\usepackage{enumitem}

\usepackage{tkz-orm}
\usetikzlibrary{arrows,shapes,automata,backgrounds,petri,decorations.markings}
\usetikzlibrary{matrix,positioning,decorations.pathreplacing,calc,tikzmark}

\tikzset{
  ln/.style  = { draw, thick, fill=black, circle, inner sep=0.3mm},
  sqloc/.style  = { draw, thick, inner sep=1.0mm },
  loc/.style    = { draw, circle, thick, inner sep=1.0mm },
  notice/.style= { draw, rectangle callout, thick, rounded corners=5pt,fill=blue!10,callout relative pointer={#1} },
  treenode/.style = {align=center, inner sep=0pt, text centered,
    font=\sffamily},
  arn_n/.style = {treenode, rectangle, font=\sffamily\bfseries },
  mymat/.style = { 
    left delimiter={[}, right delimiter ={]},nodes={anchor=base east} },
  align at top/.style={baseline=(current bounding box.north)},
  stack/.style={rectangle split, rectangle split parts=#1,draw, anchor=center}
}

\usepackage{verbatim}
	
\def\cca#1{\cellcolor{blue!#10}\ifnum #1>5\color{white}\fi{#1}}

\usepackage{tikz}
\usetikzlibrary{tikzmark,decorations.pathreplacing,arrows,shapes,positioning,shadows,trees,shapes.gates.logic.US,arrows.meta,shapes,automata,petri,calc}
\usepackage{caption}
\usepackage{adjustbox}

\usepackage[colorinlistoftodos,prependcaption,textsize=tiny]{todonotes}
\newcommandx{\unsure}[2][1=]{\todo[linecolor=red,backgroundcolor=red!25,bordercolor=red,#1]{#2}}
\newcommandx{\change}[2][1=]{\todo[linecolor=blue,backgroundcolor=blue!25,bordercolor=blue,#1]{#2}}
\newcommandx{\info}[2][1=]{\todo[linecolor=OliveGreen,backgroundcolor=OliveGreen!25,bordercolor=OliveGreen,#1]{#2}}
\newcommandx{\improve}[2][1=]{\todo[linecolor=Plum,backgroundcolor=Plum!25,bordercolor=Plum,#1]{#2}}
\usepackage{verbatim}
\usepackage[normalem]{ulem}
\usepackage{scalefnt}
\usepackage{wrapfig}
\usepackage{pgfplots}
\usepackage{tikz-qtree}
\usepackage{pgfplots}

\mathchardef\mhyphen="2D
\newcommand{\shorteq}{%
  \settowidth{\@tempdima}{-}
  \resizebox{\@tempdima}{\height}{=}%
}

\colorlet{fv}{gray!55}
\colorlet{ai}{gray!10}
\colorlet{ar}{gray!38}

\usepackage[edges]{forest}
\usepackage[T1]{fontenc}

\tikzset{%
  parent/.style={align=center,text width=3cm,rounded corners=3pt},
  child/.style={align=center,text width=3cm,rounded corners=3pt},
}

\def\addlegendimage{\csname pgfplots@addlegendimage\endcsname}

 \usepackage{tikz}
 \usepackage{xparse}
\pgfdeclarelayer{myback}
\pgfsetlayers{myback,background,main}
\usetikzlibrary{tikzmark,decorations.pathreplacing,arrows,shapes,positioning,shadows,trees,shapes.gates.logic.US,arrows.meta,shapes,automata,petri,calc,matrix,backgrounds}
\tikzset{mycolor/.style = {line width=1bp,color=#1}}%
\tikzset{myfillcolor/.style = {draw,fill=#1}}%

\NewDocumentCommand{\highlight}{O{blue!40} m m}{%
\draw[mycolor=#1] (#2.north west)rectangle (#3.south east);
}

\NewDocumentCommand{\fhighlight}{O{blue!40} m m}{%
\draw[myfillcolor=#1] (#2.north west)rectangle (#3.south east);
}
 \usetikzlibrary{matrix,decorations.pathreplacing, calc, positioning}

\newcolumntype{+}{!{\vrule width 2pt}}
\renewcommand{\thesubfigure}{\Alph{subfigure}}

\newlength\savedwidth

\usepackage{array}
\newcolumntype{L}[1]{>{\raggedright\let\newline\\\arraybackslash\hspace{0pt}}m{#1}}
\newcolumntype{C}[1]{>{\centering\let\newline\\\arraybackslash\hspace{0pt}}m{#1}}
\newcolumntype{R}[1]{>{\raggedleft\let\newline\\\arraybackslash\hspace{0pt}}m{#1}}

\renewcommand{\thesubfigure}{\Alph{subfigure}}

\newcommand{\booleans}{\mathbb{B}}
\newcommand{\naturals}{\mathbb{N}}

\newcommand{\reals}{\mathbb{R}}

\newcommand{\union}{{\cup} }

\newcommand{\nodes}{\mathcal{N}}

\newcommand*{\var}{Var}
\newcommand{\wff}{$\mathcal{W}$} 
\newcommand*{\ltf}{\mathcal{F}} 
\newcommand*{\prop}{\mathcal{P}}
\newcommand*{\op}{\mathcal{O}}
\newcommand*{\Next}{\textsf{X}}
\newcommand*{\until}{\textsf{\,U\,}}
\newcommand*{\finally}{\textsf{F}}
\newcommand*{\globally}{\textsf{G}}

\newcommand*{\sample}{\mathcal{S}}
\newcommand{\model}{m}

\newcommand{\A}{\mathcal{A}}
\newcommand{\Mm}{\mathcal{M}}

\newcommand{\varpsi}{\psi}

\newcommand{\depth}{\ensuremath{d}}
\newcommand{\discount}{\beta}
\newcommand{\arrr}{\alpha}

\newcommand{\ourtool}{\textsc{QuantLearn}}

\pgfplotsset{compat=1.8}

\usepackage[textsize=tiny]{todonotes}

\begin{document}

\title{LTL-Based Non-Markovian Inverse Reinforcement Learning}

\author{
    Mohammad Afzal\inst{1,2} \and
    Sankalp Gambhir\inst{3} \and
    Ashutosh Gupta\inst{1} \and
    Krishna S\inst{1} \and
    Ashutosh Trivedi\inst{4} \and
    Alvaro Velasquez\inst{4}
}

\institute{
    Indian Institute of Technology, Bombay, India \\
    \email{$\{$afzal,akg,krishnas$\}$@cse.iitb.ac.in}
    \and
    TCS Research, India
    \and
    École Polytechnique Fédérale de Lausanne, Switzerland \\
    \email{sankalp.gambhir@epfl.ch}
    \and
    University of Colorado, Boulder, USA \\
    \email{$\{$ashutosh.trivedi, alvaro.velasquez$\}$@colorado.edu}
}

\maketitle

\begin{abstract}
    The successes of reinforcement learning in recent years are underpinned by the characterization of suitable reward functions. However, in settings where such rewards are non-intuitive, difficult to define, or otherwise error-prone in their definition, it is useful to instead learn the reward signal from expert demonstrations. This is the crux of \emph{inverse reinforcement learning} (IRL). While eliciting learning requirements in the form of scalar reward signals has been shown to effective,  such representations lack explainability and lead to opaque learning. We aim to mitigate this situation by presenting a novel IRL method for eliciting declarative learning requirements in the form of a popular formal logic---Linear Temporal Logic (LTL)---from a set of traces given by the expert policy. A key novelty of the proposed approach is quantitative semantics of satisfaction of an LTL formula by a word that, following Occam's razor principle, incentivizes simpler explanations. Given a sample $\sample=(P,N)$ consisting of positive traces $P$ and negative traces $N$, the proposed algorithms automate the search for a formula $\varphi$ which provides the simplest explanation (in the $\globally\finally$ fragment of LTL) of the samples. We have implemented this approach as an open-source tool \ourtool{} to perform logic-based non-Markovian IRL. Our results demonstrate the feasibility of the proposed approach in eliciting intuitive LTL-based reward signals from noisy data.
    \end{abstract}

\section{Introduction}

\label{sec:intro}
Learning from demonstrations has become a viable approach to learning in environments where domain experts or performant agents can provide traces of (un-) desirable behavior. 
One important embodiment of this form of learning is known as \emph{inverse reinforcement learning}~\cite{russell1998learning} (IRL), whereby an apprentice agent learns the reward function being optimized by a given expert policy or behavior. 
IRL is indispensable in settings where it is difficult or error-prone to explicate a reward signal that captures the underlying learning objective.

\vspace{0.5em}\noindent\textbf{Interpretability and Explainability.} 
 Ng and Russell~\cite{ng2000algorithms} present a convincing argument for IRL---over apprenticeship or imitation learning---by noting that learning a reward instead of policies results in a more succinct, robust, and transferable description of the behavior.
While eliciting the objective function provides aforementioned advantages over policies, the scalar reward representations lack interpretability: \emph{Why does an action fetch different rewards in different states? What's the difference between choosing an action with a reward $4.999$ vs. another with reward $5.001$? How would the reward signal change for a more myopic (lower discount factor) agent?} 
Furthermore, the scalar reward based explanation of an objective is akin to reading a program in an assembly language (easier for machines, tedious for humans). The resulting lack of human-readable specifications (a lack of explainability) further hinders the application of verification and validation  approaches in ensuring the trustworthiness of the learning-based system design methodology.
In addition, conventionally, IRL has relied on the assumption that the reward to be learned is Markovian. However, this is a restrictive assumption that limits the space of objectives \cite{abel2021expressivity,icarte2018using}. 
\emph{We extend the interpretability and explainability of inverse reinforcement learning by eliciting logic-based objectives from demonstrations instead of scalar rewards.} 

\vspace{0.5em}\noindent\textbf{Linear Temporal Logic (LTL).} 
We focus on (a subset of) LTL~\cite{pnueli1977temporal} as the specification language due to its succinctness~\cite{BK08,gastin2001fast} and  relevance in the AI~\cite{brafman2019planning,de2013linear}, formal methods~\cite{BK08,HahnPSSTW19}, control theory~\cite{Bozkurt0ZP20,sadigh2014learning}, and machine learning~\cite{camacho2019ltl} communities.
Recently, it has gained popularity~\cite{camacho2019ltl,bozkurt2020control,hahn2021mungojerrie} in expressing learning objectives in model-free reinforcement learning (RL).


The key computational problem for the LTL-based IRL is the following: given a pair $\sample=(P,N)$ of samples consisting of positive traces $P$ and negative traces $N$ (both are sets of finite words), produce the highest-ranking LTL specification consistent with the sample where rank is informed by some user-tunable notion of simplicity over the LTL specifications.
And, here lies our central challenge of induction: we need to infer an LTL specification over unbounded length traces by observing a finite set of finite examples and counterexamples!

\vspace{0.5em}\noindent\textbf{Contributions.} 
We define quantitative semantics of satisfaction of an LTL specification over a finite word guided by a notion of parsimony of explanation. 
The complexity of an LTL can result from two aspects: the complexity of the \emph{temporal structure} ($\globally p$ is simpler than $p \wedge \Next p \wedge \Next\Next p$ in explaining the sample  $P = \{\{p\}\{p\}\{p\}\}$ and $N = \emptyset$) and the complexity of the \emph{nesting structure}  (the formula $p$ is simpler than $p \wedge \neg q$ for explaining the sample $P = \{p\}$ and $N = \emptyset$).
We expose hyperparameters (temporal discounting $\arrr$ and nesting discounting $\discount$) to take user preference in weighing these sources of complexity.
To avoid overfitting, we focus on the $\globally\finally$ fragment~\cite{DBLP:conf/cav/KretinskyE12,DBLP:conf/lics/EsparzaKS18} of LTL (temporal operators are restricted to $\globally$ and $\finally$).


We propose three optimization algorithms: \emph{Constraint system optimization} (Algorithm \ref{algo:opt}), \emph{Compositional Ranking} (Algorithm \ref{alg:composition}), and \emph{Hybrid Pattern Matching} (Algorithm \ref{algo:hybrid}), 
leveraging the state-of-the-art constraint-solver Z3~\cite{z3}, to solve the LTL formula learning problem. We show the soundness and completeness of Algorithms \ref{algo:opt} and \ref{algo:hybrid}; Algorithm \ref{alg:composition} though incomplete, is faster by 3 orders of magnitude wrt Algorithms \ref{algo:opt} and  \ref{algo:hybrid}. 
We implemented our algorithms as an open-source tool. We demonstrate the effectiveness of our method on randomized gridworld environments by computing the inverse learning error (ILE) for policies computed using the reward learned by our tool and compare it with the reward learned by a competing approach. 

\vspace{0.5em}\noindent\textbf{Organization.} We begin the technical discussion by providing a motivating example in the next section. 
In Section~\ref{sec:model} we introduce LTL and formally introduce the key problem of learning LTL from traces.
In Section~\ref{subsec:motivation}, we justify our rationale for learning a simpler form of LTL formulae, and in Section~\ref{sec:learn} provide three algorithms to learn such formulae.
We discuss our implementation and experimental results in Section~\ref{sec:experiments} and discuss related work in Section~\ref{sec:related}.
We conclude the paper by briefly discussing the contributions and suggesting some directions for potential future work.

\section{A Motivating Example}
\label{sec:motivation}
%
\newcommand{\vertic}[1]{ {\scriptsize $\texttt{#1}$} }

We assume the setting of non-Markovian reward decision processes (NMRDPs) as the model of the agent-environment interactions. Since we are interested in learning qualitative behavior, we assume a binary reward signal that captures the acceptance semantics of the underlying language.

\begin{figure}[t]
    \centering
    
  \begin{center}
    \includegraphics[width=4cm]{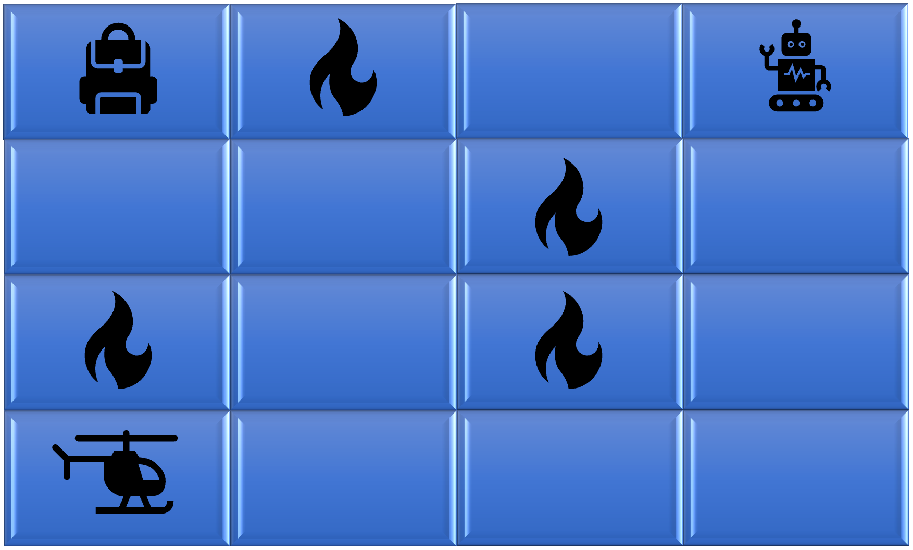}
    \hspace{3cm}
    \begin{tikzpicture}[-,>=stealth',shorten >=1pt,auto,node distance=1cm,semithick,xscale=0.2,yscale=1]
      \node[] (A)    at (0,2)             {$\vee$};
      \node[] (B)    at (-4, 1.4)           {$\land$}; 
      \node[] (B1)   at (4, 1.4)            {$\finally$};
      \node[] (C)    at (-6, 0.7)           {$\globally$}; 
      \node[] (C2)   [below=.2cm of B1]   {$\globally$}; 
      \node[] (C3)   [below=.2cm of C2]   {$q$};
      \node[] (C1)   at (-1, 0.7)            {$\finally$};
      \node[] (D)    [below=.2cm of C]    {$q$}; 
      \node[] (D1)   [below=.2cm of C1]   {$r$}; 
      
      \path (A)  edge node {} (B) ;
      \path (A)  edge node {} (B1) ; 
      \path (B)  edge node {} (C) ; 
      \path (B)  edge node {} (C1) ; 
      \path (C)  edge node {} (D) ; 
      \path (C1) edge node {} (D1) ; 
      \path (B1) edge node {} (C2) ;
      \path (C2) edge node {} (C3) ;
    \end{tikzpicture}
  \end{center} \mbox{} 
  (a) \hspace{7cm} (b)
    \caption{(a) Disaster response example (b) The syntax tree for $[\globally q \land \finally r] \vee (\finally \globally q)$}
    \label{fig:syntax}
\end{figure}

\begin{definition}[Non-Markovian Reward Decision Process]
An NMRDP is a tuple $M = (S, s_0, A, T, R, L, \Sigma)$, where $S$ is a finite set of states, $s_0 \in S$ is a distinguished initial state, $A$ is a finite set of actions, $T : S \times A \times S \to [0,1]$ is a probabilistic transition function, $\Sigma$ is an alphabet (i.e. the power set $2^\prop$ of a set of atomic propositions $\prop$), $L : S \to \Sigma$ is a labeling function, and $R : S^* \to \{0, 1\}$ is a non-Markovian reward function.
\end{definition}


The labeling function $L: S \to 2^\prop$ maps states of the NMRDP to the alphabet $\Sigma = 2^\prop$ that defines the language of the underlying objective. This alphabet denotes semantically meaningful events observed in given states. Given a sequence of states $s_i, s_j, \dots, s_k$, the corresponding trace is given by $L(s_i), L(s_j), \dots, L(s_k)$. Given positive and negative examples of such traces, the proposed non-Markovian IRL solution can learn the underlying LTL formula that captures the objective of the agent.

Given an NMRDP, the objective of reinforcement learning (RL) is to learn an optimal policy $\pi: S \times \cdots \times S \to \mathcal{D}(A)$, where $\mathcal{D}(A)$ denotes the space of probability distributions over the action space $A$. An optimal policy $\pi$ is one that maximizes the expected reward observed by the agent, as given by Equation (\ref{equation:expectedReward}), where $\tilde{s} = (\tilde{s}_0, \tilde{s}_1, \tilde{s}_2, \dots)$ is an arbitrary sequence of states sampled based on the enacted policy $\pi$ and $\gamma \in [0,1)$ is a discount factor. 
\begin{equation}
    V_\pi (s) = \lim_{n \rightarrow \infty} \mathbb{E} \left[ \sum_{t = 0}^n \gamma^t R(\tilde{s}_0, \dots, \tilde{s}_t) | \tilde{s}_0 = s \right]
    \label{equation:expectedReward}
\end{equation}
The objective is to learn a reward signal that best characterizes the given observations of behavior from an expert. For a given  process, the performance of the learned reward can be measured in terms of the inverse learning error (ILE) given by $|| V_{\pi^*_\text{true}} - V_{\pi^*_\text{learned}} ||_2$ \cite{choi2011inverse}, where $V_{\pi^*_\text{true}}$ denotes the value function computed using the optimal policy $\pi^*_\text{true}$ from the true reward signal and $V_{\pi^*_\text{learned}}$ denotes the same for the policy $\pi^*_\text{learned}$ derived from the reward signal learned using IRL. the value functions are computed over the NMRDP $M$ by leveraging the learned policy $\pi^*_\text{learned}$ using the true reward function (the one defined by $A_\text{true}$). A learned reward signal is said to generalize if the ILE is low when compared to the ground truth. 


Consider the RL problem illustrated in Figure \ref{fig:syntax}(a). The NMRDP that defines this problem is given by $M = (S, s_{14}, A, P, L, R)$, where $S$ consists of sixteen states as denoted by the grid position of the agent, with starting state $s_{14}$ (top right, at position ($1, 4$)). The action space $A$ is given by the cardinal directions with the obvious deterministic transition dynamics $P$. The labeling function is given by $L(s_{12}) = L(s_{23}) = L(s_{31}) = L(s_{33}) = \{\texttt{danger}$\}, $L(s_{11}) = \{\texttt{cargo}\}$, and $L(s_{41}) = \{\texttt{rendezvous}\}$. Suppose that the objective of the agent is to eventually obtain the cargo and, once the cargo is obtained, the agent must avoid the dangerous areas and eventually reach the rendezvous point. This is given by the LTL formula $\phi = (\finally \texttt{cargo}) \wedge [\globally (\texttt{cargo} \implies (\globally \neg \texttt{danger}) \wedge (\finally \texttt{rendezvous}))]$. In the IRL context, this objective is not known to the agent directly and it must instead learn from expert demonstrations. 

Our proposed approach enables the use of both positive and negative traces of behavior to guide the learning process. In this example, a positive trace can be $\{\texttt{danger}\}\{\texttt{cargo}\}\{\texttt{rendezvous}\}$ since it clearly satisfies the underlying objective given by $\phi$. A bad trace can be $\{\texttt{danger}\}\{\texttt{cargo}\}\{\texttt{danger}\}\{\texttt{rendezvous}\}$ as generated by the policy of moving left three times and then down three times. Since trajectories of states given by the expert (or produced by the expert policy) have a corresponding trace induced by the labeling function $L$, we assume that the traces are given over the alphabet $\Sigma = 2^\prop$ as opposed to the state space of the NMRDP.

\section{Preliminaries}
\label{sec:model}

\noindent {\bf{Propositional Logic}}. Let \emph{\var} be a set of \emph{propositional} variables, which take values from $\booleans = \{0, 1\}$ ($0$ interpreted as $false$ and $1$ as $true$). The set of formulae \wff{} in propositional logic --- with formulae denoted herein as Greek letters --- is defined inductively as follows: 
\begin{equation}
\nonumber
\varphi ::= p \in \prop~|~\varphi_1 \wedge \varphi_2~|~\neg
\varphi
\end{equation}

We use the usual syntactic sugar $\varphi \lor \varpsi$, $\varphi \Rightarrow \varpsi$, and $\varphi \Leftrightarrow \varpsi$.

A \emph{propositional valuation} is defined as a mapping $v: \var \to
\booleans$, which maps propositional variables to Boolean values. 
The semantics of this logic, given by the satisfaction relation $\models$, are
defined inductively as:
(1) $v \models x$ iff $v(x) = 1$;
(2) $v \models \neg \varphi$ iff $v \not\models \varphi$; 
  (3) $v \models \varphi \land \varpsi$ iff $v \models \varphi$ and $v \models \varpsi$.  
If $v \models \varphi$ we say $v$ models $\varphi$. A formula is said to be
\emph{satisfiable} if there exists a model for it.
There are practical tools, called SAT solvers, that can check satisfiability of the formulae.
\smallskip 

An \emph{alphabet} $\Sigma$ is a non-empty,
finite set of \emph{symbols}. A \emph{finite word} $w$ over $\Sigma$ is a finite
sequence $a_1 a_2 ... a_n$ of symbols from $\Sigma$. 
The empty sequence is called the empty word, denoted $\epsilon$. The domain of $w$, denoted $\mathsf{dom}(w)$ is the set of positions in $w$. Thus, $\mathsf{dom}(a_1 \dots a_n)=\{1,2,\dots,n\}$ and $\mathsf{dom}(\epsilon)=\emptyset$. The length of a finite word $w$ is denoted $| w|$, with $| \epsilon | = 0$. The set of all finite words over $\Sigma$ is denoted $\Sigma^*$.
An \emph{infinite word} over $\Sigma$ is an infinite sequence $w = a_0 a_1
a_2 ...$ of symbols $a_i \in \Sigma$ with $i \in \naturals$. The set of all
infinite words over $\Sigma$ is denoted $\Sigma^\omega$. 

Given a finite word $w$, we write $w(i)$ for the symbol at position $i$.
The subsequence of $w$ from index $i$ to index $j$, both inclusive, is
denoted as $w[i:j]$, while $w[i:]$ denotes the suffix of $w$ starting from index $i$.
When clear from context, we write $w_i$ for $w[i:]$.

\vspace{0.5em}\noindent\textbf{Linear Temporal Logic (LTL).} Let $\prop$ be
a set of propositional variables. LTL \cite{DBLP:books/daglib/0020348} is an
extension of propositional logic with temporal modalities, which allows the expression of temporal properties. Formulae in LTL are defined inductively. An LTL formula $\varphi$ over $\prop$ is defined by the following grammar:
\begin{equation}
\nonumber
\varphi ::= true ~|~a \in \prop~|~\varphi_1 \wedge \varphi_2~|~\neg
\varphi~|~\Next~\varphi~|~\varphi \until \varphi
\end{equation}

      

Using the above, the formulae $\finally \varphi=true \until
\varphi$ and $\globally \varphi=\neg \finally \:\neg \varphi$ can be derived. The size of an
LTL formula $\varphi$ denoted by $|\varphi|$ is the number of subformulae in
it. For example, if $\varphi=p \until \psi$, then $|\varphi|=|\psi|+2$. 
We say that an infinite word $\alpha$ satisfies an LTL formula $\phi$, and we write $\alpha \models \phi$, if:
\begin{itemize}
    \item $\alpha \models a \text{ iff } a \in \alpha[0:0]$
    \item $\alpha \models \neg \Phi \text{ iff } \alpha \not\models \Phi$
    \item $\alpha \models \Phi \land \Psi \text{ iff } \alpha \models \Phi \text{ and } \alpha \models \Psi$
    \item $\alpha \models \Next \Phi \text{ iff } \alpha[1:] \models \Phi $
    \item $\alpha \models \Phi \until \Psi \text{ iff } \exists~i \text{ s.t. } \alpha[i:] \models \Psi \text{, and } \forall j <i, \alpha[j:] \models \Phi$.
\end{itemize}

%
The language $L(\varphi)$ of an LTL formula $\varphi$ is defined as
\begin{equation*}
L(\varphi) = \{\alpha \in \Sigma^* | \alpha \models \varphi\}~.
\end{equation*}
Two LTL formulae having the
same language are called {\em equivalent}. In this paper, we learn formulae that
are in the $\globally\finally$-fragment of LTL, where only the $\globally$ and
$\finally$ modalities are allowed apart from Boolean connectives.
Since the LTL formulae can be converted into negation normal form (NNF), we
learn formulae only in NNF. 
Given an LTL formula $\varphi$, the syntax tree of $\varphi$ is a tree labeled
 with variables, Boolean connectives and temporal modalities.
 The variables always appear at the leaf nodes, while temporal
 modalities and Boolean connectives are internal nodes.
 For example, we present the
  syntax tree of the formula $\varphi=[\globally q \land \finally r] \vee
(\finally \globally q)$ in Figure \ref{fig:syntax}(b). 
The depth of an LTL formula is the maximum of distances of the root to leaves in the corresponding syntax tree. The depth in Figure \ref{fig:syntax}(b) is 3.
Although, LTL is defined over infinite words, we observe only finite executions of the systems. Therefore, we defined samples with finite words as follows.

\vspace{0.5em}\noindent\textbf{Samples.} A \emph{sample} is a pair $\sample=(P, N)$ of 
two finite, disjoint sets $P, N \subseteq (2^{\prop})^*$.
The words in $P$ are \emph{positive traces} while words in $N$ are
\emph{negative traces}. For an LTL formula \(\varphi\), we say \(\sample 
\models \varphi\) iff \(\forall \tau \in P,\; \tau \models \varphi\) 
and \(\forall \tau' \in N,\; \tau' \not \models \varphi\). 
We learn an LTL formula from a given sample.


\section{Occam's Razor for LTL} 
\label{subsec:motivation}
Given an LTL formula $\varphi$  and a finite word $w$, we design a valuation function  $V(\varphi, w)$ that quantifies the parsimony of $\varphi$ in explaining $w$. 
Intuitively, a pair scores high if all of the subformulae of the formula $\varphi$ contribute in
accepting $w$ in $L(\varphi)$. 
However, we do so in a nuanced fashion by geometrically attenuating the effect of parsimony with the length of the word.
For example, $\globally(p\lor q)$ should score
well along with word $(\{p\}\{q\})^3$ but should not do well
with word $(\{p\})^6$, since the subformula $q$ did not contribute to the acceptance. 
Similarly, $\globally(p\lor q)$ should score
better along with word $(\{p\}\{q\})^3$ than $(\{p\})^5\{q\}$. 

\subsection{Quantifying Expressive Parsimony}
Let us present a valuation function first.
Let $\ltf$ represent the set of NNF $\globally\finally$-fragment formulae over $\prop$.  
We interpret LTL formulae over finite words and define the quantitative
semantics in terms of a \emph{valuation mapping} $V : \ltf \times \Sigma^* \to
\reals ^+ \: \union \: \{0\}$, where $\Sigma=2^{\prop}$. The valuation mapping
is defined over a word $w \in \Sigma^*$ inductively:
\begin{eqnarray*}
V(p, w) \!\!\! &=& \!\!\!  \begin{cases} 
    1 & \textnormal{if } p \in w(1) \\
    0 & \textnormal{otherwise}
   \end{cases} \\
V(\lnot p, w) \!\!\! &=& \!\!\!  
    \begin{cases} 
    1 & \textnormal{if } p \not\in w(1)\\
    0 & \textnormal{otherwise}
    \end{cases} \\
V(\varphi {\land} \psi, w) \!\!\! &=& \!\!\! \discount \cdot V(\varphi, w) \cdot V(\psi, w) \\ 
V(\varphi {\lor} \psi, w) \!\!\! &=& \!\!\! \discount  \cdot \dfrac{V(\varphi, w) + V(\psi, w)}{2} \\ 
 V(\globally \varphi, w) \!\!\!\! &=& \!\!\!\!
    \begin{cases}
    \discount \sum\limits_{i = 0}^{|w|} \arrr^{i} V(\varphi, w_i) & \! 
     \text{ if } V(\neg\varphi, w_t) {=} 0, \text{ $\forall t$}\\ 
         0 & \! \textnormal{otherwise}
     \end{cases} \\ 
V(\finally \varphi, w) \!\!\!\! &=& \!\!\!\! 
\begin{cases}
\discount \arrr^{t} V(\varphi, w_t) & \! t {=} \min \{j \mid V(\varphi, w_j) {>} 0\} \\ 
         0 & \! \textnormal{if } V(\varphi, w_t) = 0, \text{ $ \forall t$}
\end{cases}
\end{eqnarray*}
Here $w_j$ is a shorthand for $w[j:]$.
If $w \models \varphi$, then $V(\varphi,w)$ is non-zero.
This scheme is parameterized by two discount factors: the \emph{temporal discount factor} $\arrr$ and the \emph{nesting discount factor}  $\discount$.

For the literals, we assign valuation zero or one if the word satisfies
the literals or not.
We interpret conjunction as multiplication, which implies we need both subformulae
to do well on the word.
We interpret disjunction as an addition, which implies we give a high score to the formula
if any of the two subformulae does well on the word.
Our interpretation of $\globally \varphi$ computes the discounted sum of the value of $\varphi$
at each position of the word.
To reduce the weight of a letter appearing further in the word, we apply the temporal discount of $0 < \arrr < 1$.
Our interpretation of $\finally \varphi$ computes the discounted score of $\varphi$
at the earliest position where $\varphi$ has a non-zero score.
We apply the nesting $0 < \discount < 1$ each time we construct a more complex formula, i.e. we go deeper in the nested structure of the LTL formula.
The discount values $\discount$ and $\arrr$ are parameters defined by the user to control nesting complexity and temporal complexity.
\begin{example}
Consider the
formula $\varphi=\finally q$ for $p, q \in \prop$ and the word
$w=\{p\}\{p,r\}\{p\}\{p,s\}\{p\}\{p\}\{p,q\}(\{r\}\{q\})^*$. Then $q$ holds for
the first time at the position $t=7$, and for all $t' < t$, $q$ is not present
in $w$. Thus, $V(q,w[7:])=1$, making $V(\varphi, w)=\discount\arrr^{7}$.
 Note that our valuation assigns non-zero scores only for satisfiable formulas. 
\end{example}

We extend the notion of a valuation from a word to a sample in a natural fashion.
For a sample $\sample$, the valuation of $\varphi$ is taken as the sum of valuations over all positive traces in the sample:
\[
V(\varphi, \sample = (P, N)) = \sum_{w \in P}^{} V(\varphi, w).
\]
This scheme attempts to match intuition about the operators. We posit that while the definition of the valuation functions is subjective, natural variations do not provide any significant theoretical advantages to our paradigm. For example, One may assign $V(\varphi {\land} \varpsi, w)$ as the minimum of $V(\varphi, w)$ and $V(\varpsi, w)$~\cite{tabuada2015robust}. This valuation ensures that both $\varphi$ and $\varpsi$ must score high for $\varphi \land \varpsi$ to score high. However, the function is not sensitive to the formula that has a higher value. Therefore, the learning algorithm becomes unguided for one part of the formula. This suggests a modification to the valuation function that takes both the subformulae into account symmetrically, without flattening one of the subformulae. Our valuation function for the conjunction of two formulae, defined as their product $V(\varphi, w) \cdot V(\varpsi, w)$ is based on this idea.

\section{Learning Algorithms}
\label{sec:learn}
\label{sec:algo}

As our main contribution, we propose learning algorithms to solve the following
problem.  
\emph{Given a sample $\sample$=(P,N) over finite words, compute an LTL formula  
$\varphi$ in the $\globally\finally$-fragment that best describes $\sample$ and
is consistent with $\sample$. That is, $\varphi$ has the highest score,  based
on the valuation described above,  
among all formulae such that for all $w \in P$, $w \models \varphi$ and for all
$w' \in N$, $w' \not\models \varphi$}.



To achieve the goal of ranking formulae based on a quantitative notion of satisfiability, we propose the techniques of
\textit{Constraint System Optimization} (Section \ref{sec:cso})
\textit{Optimized Pattern Matching} (Section \ref{sec:opm}) and \emph{Hybrid
Pattern Matching} (Section \ref{sec:hybrid}). In the first one, we get a
sample and a depth $d$ as input. We encode the syntax tree of this unknown
formula of depth $d$ along with constraints to compute the score of each node
in the tree.
The second one makes use of a formula template pattern  
provided by the user, but has unknown propositional variables. We encode
constraints which allow mapping these variables to unique variables occurring in
the sample. The third one is a ``hybrid'' approach, a middle ground
incorporating both of the above techniques. 
We use an
optimizing SMT solver to solve the constraints to find 
the \emph{best} formula for the sample with the highest score according to the
valuation function.

 


 
 \begin{algorithm}[t]
    \begin{algorithmic}[1]
        \Procedure{ConstraintOpt}{$\sample = (P, N), d$} 
        
            \State construct $\Phi^\sample_\textnormal{d}$ \Comment{\textcolor{red!50!black}{Constraints in eq(\ref{eq:constr})}}
            \State max.
            $\min( \{y_{1,0}^\tau\: |\: \tau \in P\} )$ with
            $\Phi^\sample_\textnormal{d}$ as a constraint
            
            \If{optimization succeeds with model m} \Comment SAT \State
                construct formula tree from m \State \textbf{return}
                optimized formula tree \Else \State \textbf{return} UNSAT \EndIf
                \EndProcedure
    \end{algorithmic}
    \caption{Computing the optimal formula given a sample}
    \label{algo:opt}
\end{algorithm}

 \subsection{Constraint System Optimization}
 \label{sec:cso}
 
In Algorithm \ref{algo:opt}, we present the method to compute the optimal
formula with the highest score for a sample $\mathcal{S} = (P,N)$ along with the
desired depth $\depth$ of the formula. This is obtained by computing and optimizing
the scores of a class of formulae of depth $\depth$,  constrained to be well formed
and to be satisfied by $\mathcal{S}$. We reduce
the construction of an LTL formula for a sample $\sample$ to a constraint system
$\Phi ^\sample _\depth$, which is constructed in three parts as:
\begin{align}\label{eq:constr}
  \textstyle
    \Phi ^\sample _\depth = \varphi^\textnormal{ST} _\depth \land \bigwedge_{\tau \in P} \varphi^\tau _\depth 
    \land \bigwedge_{\tau' \in N} \neg \varphi^{\tau'} _\depth
\end{align}
The first part encodes the structure of a \emph{syntax tree} (ST)
representing an unknown formula, while the second and third encode the
functional constraints on the \emph{score} of each node enforced by
the operators that make up the formula. 

To encode the formula structure, we use a syntax tree with  identifiers
$\nodes=\{1, 2, \ldots, n\}$ for the set of nodes, where $n = |\nodes|$.
We assume herein that the root node is
identified as $1$.  We have the child relations $L$ and $R$, such that $(i, j)
\in L$ (or $R$) iff the node $j$ is the left (right) child of node $i$. The only child of unary operators  
is considered as a  left child by assumption. For each node $i \in \nodes$ and
possible label $\lambda \in \op \;\union\; \prop$, we introduce a Boolean variable
$x_{i, \lambda}$ indicating whether the node is labelled with an operator
($\op$) or variable ($\prop$).
The formula $\varphi^\textnormal{ST} _n$ is constructed as the conjunction of formulae (\ref{eq:node_type}) through (\ref{eq:finally_score}).
\begin{eqnarray}
   \hspace*{-12pt} \left [\bigwedge\limits_{1 \leq i \leq n} \bigvee\limits_{\lambda \in \prop \union \op} \hspace{0em} x_{i, \lambda} \right] \hspace{0em}&\land& \hspace{0em}\left [\bigwedge\limits_{1 \leq i \leq n} \bigwedge\limits_{\substack{\lambda \not = \lambda' \\\in \prop \union \op}} (\neg x_{i, \lambda} \lor \neg x_{i, \lambda'} )\right] \label{eq:node_type} \\
  \hspace*{-12pt} \bigwedge\limits_{\not \exists j (i,j) \in L} \; \bigvee\limits_{p \in \prop} x_{i, p} && \label{eq:leaf}
\end{eqnarray}
The constraint given by expression \eqref{eq:node_type} ensures the two
properties that each node must accept at least one label, and that it must accept
at most one label, while expression \eqref{eq:leaf} ensures that the leaf nodes
are labelled with propositional variables, and not operators (since they have no children).  

Next, we encode the functional constraints imposed by the operators. For this,
 to each node $i \in \nodes$, we attach a set of real variables $Y^\tau _i = \{y^\tau _{i, t} \; \mid 0 \leq t \leq |\tau|, i \in \nodes\}$ representing its
 \emph{score} at each point $0 \leq t \leq |\tau|$ in a trace $\tau \in  P \cup N$, where $y^\tau _{i, t}$ is defined below.  For a given trace $\tau$ we construct
$\varphi^\tau _\depth$ as the conjunction of:
\begin{gather}
  y^\tau _{1, 0} > 0 \label{eq:pos_def} \\
  \textstyle
    \bigwedge\limits_{1 \leq i \leq n} \bigwedge\limits_{p \in \prop}\;\;x_{i, p} \Rightarrow \left[ 
        \bigwedge\limits _{1 \leq t \leq |\tau|} y^\tau _{i, t} = \begin{cases}
            1 & \textnormal{if } p \in \tau (t) \\
            0 & \textnormal{if } p \not \in \tau (t)
        \end{cases}
      \right] \label{eq:prop_score} \\
    \bigwedge \limits_{\substack{1 \leq i \leq n, \\(i, j) \in L}} \;\;x_{i, \neg} \Rightarrow \left[ 
        \bigwedge \limits_{1 \leq t \leq |\tau|} y^\tau _{i, t} = \discount\cdot\max(0, 1 {-} y^\tau _{j, t})
      \right] \label{eq:not_score} 
\end{gather}

\begin{gather}
    \bigwedge\limits _{\substack{1 \leq i \leq n, \\(i, j) \in L, \\(i, j') \in R}}\;\;x_{i, \land} \Rightarrow \left[ 
        \bigwedge\limits _{1 \leq t \leq |\tau|} y^\tau _{i, t} = \discount\cdot y^\tau _{j, t} \cdot y^\tau _{j', t}
      \right] \label{eq:and_score} \\
    \bigwedge\limits _{\substack{1 \leq i \leq n,\\ (i, j) \in L, \\(i, j') \in R}}\;\;x_{i, \lor} \Rightarrow \left[ 
        \bigwedge\limits _{1 \leq t \leq |\tau|} y^\tau _{i, t} = \discount\cdot\frac{y^\tau _{j, t} + y^\tau _{j', t}}{2}
      \right] \label{eq:or_score} \\
      \textstyle
    \bigwedge\limits _{\hspace{-4mm}\substack{\vspace{2mm}\\ 1 \leq i \leq n \\ (i, j) \in L}} x_{i, \globally} \Rightarrow 
    \!\!\!\bigwedge\limits _{1 \leq t \leq |\tau|}\left[\left[ 
         y^\tau _{i, t} = \discount\sum\limits _{\substack{t \leq t' < |\tau|}}
                \!\!\! \arrr^{t' - t} \cdot y^\tau _{j, t'}
                \right]  \land  \nonumber\right. \\
                \left.\left[ \bigwedge\limits_{t \leq t' < |\tau|} \!\!\! (y^\tau _{j, t'} > 0)\right]\right] \lor y^\tau_{i, t} = 0\label{eq:globally_score} \\
    \bigwedge\limits _{\substack{1 \leq i \leq n, (i, j) \in L}}\;\; \!\!\!\!\! x_{i, \finally} \Rightarrow\qquad \qquad \qquad \qquad \qquad \qquad \qquad  \nonumber\\
    \left[ 
        \bigwedge\limits _{1 \leq t \leq |\tau|} \exists\: t'.\: \left[ y^\tau _{i, t} = 
          \discount\cdot \arrr^{t' - t} \cdot y^\tau _{j, t'}\right] \land (t \leq t')\land \right. \nonumber\\
                \qquad\qquad\left. \left[ \bigwedge\limits_{t < t'' < t'} 
                \neg(y^\tau _{j, t''} > 0)\right] \land (y^\tau _{j, t'} > 0)\label{eq:finally_score} \right]
\end{gather}

Corresponding to Boolean variables and to each operator, the constraints encode
the calculation of the valuation of a given node as a function of the valuation
of its children, as defined in Section \ref{subsec:motivation}. 
The score for a node labelled $\globally$ at a position $t$ in the trace is
described by the constraint (\ref{eq:globally_score}). The first conjunct
encodes the actual score as a function of its child, adding the child's score
over all positions in the input word, scaled by an exponential. The second
conjunct simply ensures that the $\globally$-property holds in a classical
sense, i.e., that its child has positive valuation at all positions. Similarly,
$\finally$ is encoded in constraint (\ref{eq:finally_score}), where we look for
the first position $t'$ where its child has positive valuation. The score of its
child is exponentially scaled so as to diminish the contribution from an
occurrence far away from the start.
 
Finally, we optimize the score of the syntax tree using the score of the
root node  $y^\tau _{1, 0}$ w.r.t. the constraint system $\Phi ^\sample _\depth$ (
Algorithm \ref{algo:opt}). Then we use the resulting model to label the tree,
obtaining the optimal formula of chosen depth. By iterating over $\depth$, we may
obtain the minimal such formula.
Next, we establish the correctness and soundness of this algorithm.
\begin{lemma}[Well-Formedness]
    Any satisfying assignment of the constraint system
    (\ref{eq:constr}-\ref{eq:finally_score}) encodes a well-formed LTL formula, and
    every well-formed formula of depth \(\leq \depth\) can be encoded within
    \(\Phi^\sample_\depth\).
    \label{lem:one}
\end{lemma}

The proof is a straight forward argument. In one direction,  the constraints ensure that the 
 $\{x_{i, \lambda} \mid  \lambda \in \prop \cup \op\}$ forms a well formed binary tree. The encoding 
 requires matching this binary tree with the parse tree of the given formula. In the converse direction, 
one can recursively construct the parse  tree of a formula which satisfies the constraint system. 

    \begin{theorem}[Completeness]
        Given a sample \(\sample = (P, N)\), if there exists an LTL formula
        \(\varphi\) of depth \(\depth\) such that \(\sample \models \varphi\),
        then there exists a model \(\model\) satisfying the constraint system
        (\ref{eq:constr} - \ref{eq:finally_score}), i.e. \(\model \models
        \Phi^\sample_\depth\) such that the parse tree \((x_i)\) encodes the
        formula \(\varphi\) and for each node \(i\) and position \(t\) in each
        trace \(\tau\),  \(y_{i, t} = V(\varphi_i, \tau[t:])\), where
        \(\varphi_i\) is the formula encoded by the subtree with root at node
        \(i\).

        \label{theorem:completeness}
    \end{theorem}

    \begin{proof}
        We induct on the depth \(\depth\) of the formula. 

        \underline{Base case: \(\depth = 0\)}\\
        \(\varphi = p\) for some propositional variable \(p\), and \(\sample
        \models \varphi\).

        Setting \(x_{1, p} = 1\) in \(\model\) and resolving the constraints,
        for each positive trace \(\tau\) we are left with

        \begin{gather}
            \bigwedge_{\tau\in P} y_{1, 0}^\tau > 0 \label{eq:constraintpositivity}\\
            \bigwedge_{\tau\in P} \bigwedge_t y_{1, t}^\tau = \begin{cases}
                1 \text{ if } p \in \tau(t) \\
                0 \text{ otherwise}
            \end{cases} \label{eq:propmodelcases}
        \end{gather}

        Constraint \ref{eq:propmodelcases} encodes the valuation function \(V(p,
        \tau[t:])\). We assign the same value to it in \(\model\).

        Since for each positive trace \(\tau, \tau \models \varphi\), we have
        \(p \in \tau(0)\), and hence, \(y_{1, 0}^\tau = 1 > 0\), and the
        constraint system is satisfied.

        Similarly for negative traces, constraint \ref{eq:constraintpositivity}
        changes to \(y_{1, 0}^\tau = 0\), and since \(\tau \not \models
        \varphi\), we have \(p \not \in \tau(0)\), and the constraint system is
        satisfied.

        \uline{Induction: Given the system is complete for all depths \(\leq
         (d-1)\), we show it is complete for depth \(d\)}.

        We prove this for each top-level operator:

        \uline{Assuming \(\varphi = \psi \lor \chi\)} --- since \(\sample
        \models \varphi\), for each positive trace \(\tau\), we must have
        atleast one of \(\tau \models \psi\) and \(\tau \models \chi\), and for
        each negative trace \(\tau'\), we have \(\tau' \not \models \psi,
        \chi\).

        Construct new samples \(\sample_1 = (P_1, N)\) and \(\sample_2 = (P_2,
        N)\) with \(P_1 \subseteq P\) containing the positive traces which
        \(\psi\) satisfies, and \(P_2\) for \(\chi\). By the induction
        hypothesis, we can find satisfying models for the constraint systems
        \(\Phi^{\sample_1}_{\depth-1}\) and \(\Phi^{\sample_2}_{\depth-1}\), say
        \(\model_1\) and \(\model_2\).

        Using these, we construct a satisfying model for
        \(\Phi^{\sample}_{\depth}\), \(\model\). Assign to the \(\{x_i\}\) the
        syntax tree encoding for \(\varphi\). Through \(\model_1\), the values
        \(\{y_i\}\) of \(\psi\) are known over \(\sample_1\). Without loss of
        generality, set the unknown values (over \(P\setminus P_1\)) to 0.
        Similarly for \(\chi\) and \(P_2\).

        The constraint system for positive traces reduces again to
        
        \begin{gather}
            \bigwedge_{\tau\in P} y_{1, 0}^\tau > 0 \\
            \bigwedge_{\tau\in P} \bigwedge_t y_{1, t}^\tau = \beta\cdot\frac{y_{m, t}^\tau + y_{m', t}^\tau}{2}
        \end{gather}

        where nodes \(1, m, m'\) are the root node and its left and right
        children respectively. The models \(\model_1\) and \(\model_2\) ensure
        that both the terms on the RHS for \(y_{1, 0}^\tau\) are non-negative,
        and for each trace, atleast one is positive. Assigning the respective
        values to \(y_{n, t}^\tau\) in \(\model\), since the sum of a
        non-negative and a positive value is greater than 0, the system is
        satisfied. 

        For negative traces, we require \(y_{1, 0}^\tau = 0\). Clearly, the
        models of the subformulae ensure that both terms on the RHS for \(y_{1,
        0}^\tau\) are zero. Hence, their sum is zero as well. Hence, the
        negative trace constraints are also satisfied under this \(\model\).
        
        \(\model\) is the required model.

        \uline{Assuming \(\varphi = \psi \land \chi\)} --- we must have \(\tau
        \models \psi, \chi\), and for each negative trace \(\tau'\), we have
        atleast one of \(\tau' \not \models \psi\) and \(\tau' \not \models
        \chi\).

        The proof proceeds identically to the previous case, except with a
        partitioning of the negative traces. Similar to the sum-based constraint
        system for \(\lor\), we have the product-based system for positive
        traces with the same notation as before
        
        \begin{gather}
            \bigwedge_{\tau\in P} y_{1, 0}^\tau > 0 \\
            \bigwedge_{\tau\in P} \bigwedge_t y_{1, t}^\tau = \beta\cdot y_{m, t}^\tau \cdot y_{m', t}^\tau~.
        \end{gather}

        Again, the models of the subformulae ensure that both terms on the RHS
        for \(y_{1, 0}^\tau\) are positive, hence so is their product.

        For negative traces, without loss of generality we assign \(\mathbf{1}\)
        to the unknown values over the complements \((N\setminus N_1, N\setminus
        N_2)\). We have
        
        \begin{gather}
            \bigwedge_{\tau\in N} y_{1, 0}^\tau = 0 \\
            \bigwedge_{\tau\in N} \bigwedge_t y_{1, t}^\tau = \beta\cdot y_{m, t}^\tau \cdot y_{m', t}^\tau~,
        \end{gather}

        and the models ensure atleast one of the terms in the RHS for \(y_{1,
        0}^{\tau'}\) is zero, and hence so is the product.
        
        The constraint system is satisfied, and \(\model\) is the required
        model.

        \uline{Assuming \(\varphi = \globally\psi\)} --- we must have \(\psi\)
        satisfied at every point on every positive trace and to not be satisfied
        on some point on each negative trace, i.e.

        \begin{gather}
            \sample \models \varphi = \globally\psi \\
            \forall \tau \in P\; \forall t < |\tau|,\; \tau[t:] \models \psi~, \text{ and} \\
            \forall \tau \in N\; \exists t < |\tau|,\; \tau[t:] \not \models \psi~.
        \end{gather}

        By taking all possible suffixes of traces in \(P\), construct \(P^*\),
        and by resolving the existential quantifier, collect relevant
        (non-satisfying) suffixes of \(N\) as \(N^*\). Generate the new sample
        \(\sample^* = (P^*, N^*)\).

        By construction, \(\sample^* \models \psi\). Using the induction
        hypothesis, there exists a model \(\model^*\) satisfying the constraint
        system \(\Phi^{\sample^*}_{\depth-1}\). We use \(\model^*\) to construct
        a model \(\model\) for \(\Phi^\sample_\depth\).

        Setting the values for \(\{x_i\}\) as encoding \(\varphi\), we are left
        with the reduced constraint system

        \begin{gather}
            \bigwedge_{\tau\in P} y_{1, 0}^\tau > 0~, 
            \bigwedge_{\tau'\in N} y_{1, 0}^{\tau'} = 0 \\
        \end{gather}
        \begin{gather*}
        \bigwedge_{\tau \in P\cup N} \bigwedge _{1 \leq t \leq |\tau|} 
        \end{gather*}
        \begin{gather}
        \left[\left[ 
        y^\tau _{1, t} = 
                \beta\cdot\sum _{\substack{t \leq t' < |\tau|}}
                \alpha^{t' - t} \cdot y^\tau _{m, t'}
                \right] \land \left[ \bigwedge_{t \leq t' < |\tau|} 
                  (y^\tau _{m, t'} > 0)\right]\right]
            \lor y_{1, t}^\tau = 0~.
        \end{gather}

        For positive traces, from \(\model^*\), we have an assignment for each
        of the terms in the sum for \(y_{1, 0}^\tau\) being positive (we posed
        them as positive traces in \(P^*\)). Hence, their sum is positive too,
        and adding these inferred assignments for \(y_{1, t}\), the constraints
        are satisfied.

        For the negative traces, adding the assignments from \(\model^*\),
        clearly the constraints \(y_{m, t} > 0\) cannot be satisfied for each
        parameter in the sum for \(y_{1, 0}\), by construction. Thus, \(y_{1, 0}
        = 0\).

        \(\model\) is the required model.

        \uline{Assuming \(\varphi = \finally\psi\)} --- the proof proceeds
        identically to the previous case, with existential quantification in the
        positive case, and suffix construction in the negative, since these are
        dual operators.

        The induction is complete.

    \end{proof}

    \begin{theorem}[Soundness]
        Given a sample \(\sample\), if there exists a model \(\model\)
        satisfying the constraint system (\ref{eq:constr} - \ref{eq:finally_score}), i.e.
        \(\model \models \Phi^\sample_\depth\), then there exists an LTL formula
        \(\varphi\) of depth \(\depth\) such that the parse tree \((x_i)\)
        encodes the formula \(\varphi\) and for each node \(i\) and position
        \(t\) in each trace \(\tau\),  \(y_{i, t} = V(\varphi_i, \tau[t:])\),
        where \(\varphi_i\) is the formula encoded by the subtree with root at
        node \(i\).

        \label{theorem:soundness}
    \end{theorem}

    \begin{proof}
        We induct on the depth parameter \(\depth\) of the constraint system.

        \uline{Base case: d = 0.}

        Since there is only a single leaf node, the model must have one of the
        propositional variable mapped to it, say \(p\). With this, the
        constraint system reduces to

        \begin{gather}
            \bigwedge_{\tau\in P} y_{1, 0}^\tau > 0~, \bigwedge_{\tau\in N} y_{1, 0}^\tau = 0\\
            \bigwedge_{\tau\in P\cup N} \bigwedge_t y_{1, t}^\tau = \begin{cases}
                1 \text{ if } p \in \tau(t) \\
                0 \text{ otherwise}
            \end{cases}
        \end{gather}        

        Clearly, due to the constraints, for each trace \(y_{1, t}^\tau = V(p,
        \tau[t:])\). In particular, \(y_{1, 0}^\tau = 1\) for \(\tau \in P\),
        and with the constraints, we must have \(p \in \tau(0)\), which, by the
        LTL semantics implies \(\tau \models p\). 

        Similarly for negative traces \(\tau' \in N\) we find \(\tau'
        \not\models p\).

        Combining these, we have \(\sample \models p\), and hence \(\varphi =
        p\) is the required LTL formula of depth 0.

        \uline{Induction: Given the system is sound for all depths \(\leq
         (d-1)\), we show it is sound for depth \(d\)}.

        We prove this for each top-level operator in the constraint encoding:
        
        The top-level operator is obtained by checking which of \(\{x_{1, O}\}\)
        holds under \(\model\). The constraint system ensures there is exactly
        one. We assume the \(\{x_{i, O}\}\) have been processed to obtain
        the encoded formula, say \(\varphi\).

        Our task is now to show that \(\sample \models \varphi\).

        \uline{Assuming \(\varphi = \psi \lor \chi\)} --- We have the reduced
        constraints

        \begin{gather}
            \bigwedge_{\tau\in P} y_{1, 0}^\tau > 0~, \bigwedge_{\tau'\in N} y_{1, 0}^{\tau'} = 0 \\
            \bigwedge_{\tau\in P\cup N} \bigwedge_t y_{1, t}^\tau = \beta\cdot\frac{y_{m, t}^\tau + y_{m', t}^\tau}{2}
        \end{gather}

        For positive traces, in particular the constraints on the root-node
        \(y_{1, 0}^\tau\) imply at least one of \(y_{m, 0}^\tau > 0\) and
        \(y_{m', 0}^\tau > 0\) must hold, where \(m, m'\) are the left and right
        children of the root-node 1 respectively. And for negative traces we
        have \(y_{m, 0}^{\tau'} = y_{m', 0}^{\tau'} = 0\).
        
        In conjunction with the rest of the constraints, this is precisely two
        copies of the constraint system \(\Phi^\sample_{\depth-1}\).
        
        In either case, i.e. for the left or right child being non-zero, the
        induction hypothesis implies that the subformula holds over the sample
        \(\sample\), i.e. \(\sample \models \varphi_m\) (or \(\varphi_m'\)). And
        since this is a subformula of \(\varphi\), by introduction of \(\lor\)
        , \(\sample \models \varphi\).

        \uline{Assuming \(\varphi = \psi \land \chi\)} --- Proof proceeds
        similar to the previous case, with a case-bifurcation in the negative
        traces instead of the positive ones.

        \uline{Assuming \(\varphi = \globally\psi\)} --- Again, we have the
        reduced constraint system

        \begin{gather}
            \bigwedge_{\tau\in P} y_{1, 0}^\tau > 0~, \bigwedge_{\tau'\in N} y_{1, 0}^{\tau'} = 0 
        \end{gather}
        \begin{gather*}
        \bigwedge_{\tau \in P\cup N} \bigwedge _{1 \leq t \leq |\tau|} 
        \end{gather*}
        \begin{gather}
        \left[\left[ 
        y^\tau _{1, t} = 
                \beta\cdot\sum _{\substack{t \leq t' < |\tau|}}
                \alpha^{t' - t} \cdot y^\tau _{m, t'}
                \right]\land  \left[ \bigwedge_{t \leq t' < |\tau|} 
                  (y^\tau _{m, t'} > 0)\right]\right]
            \lor y_{1, t}^\tau = 0~,
        \end{gather}

        First, in the case of the positive traces, since \(y_{1, 0}^\tau > 0\),
        we must have the first clause of the score constraint holding (since the
        second clause \(y_{1, 0}^\tau = 0\) evaluates to false). Thus, for each
        point \(t\) in the trace, we must have \({y_{m, t} > 0}\) with notation
        as before. 

        For negative traces, we can reduce the constraints on \(y_{1, 0}^\tau\)
        further to obtain

        \begin{gather}
            \bigwedge_{\tau\in N}\bigvee_t y_{m, t} = 0~,
        \end{gather}

        implying there exists a position for which \(y_{m, t} = 0\).

        As before, taking all possible suffixes of traces in \(P\), construct
        \(P^*\), and by resolving the existential quantifier above, collect
        relevant (non-satisfying) suffixes of \(N\) as \(N^*\). Generate the new
        sample \(\sample^* = (P^*, N^*)\). Combined with the constraints above,
        we have a satisfying assignment to \(\Phi^{S^*}_{\depth-1}\) by
        restricting \(\model\) as needed to the smaller system.

        Thus, by the induction hypothesis, for each trace \(\tau \in P\),
        \(\psi\) holds on every suffix of \(\tau\). By the semantics of the
        operator \(\globally\), we have \(\tau \models \globally \psi =
        \varphi\).

        Further, for each negative trace \(\tau' \in N\), \(\psi\) does not hold
        on some suffix of \(\tau\). Again, by the semantics of \(\globally\),
        \(\tau \not \models \globally \psi = \varphi\).

        Thus, \(\sample \models \varphi\) as required.

        \uline{Assuming \(\varphi = \finally\psi\)} --- Proof proceeds similar
        to the previous case, with existential quantification on the positive
        traces instead of the negative.

        The induction is complete.

    \end{proof}

    \begin{corollary}[Valuation Equivalence]
        The valuation semantics are equivalent to the LTL semantics. For any
        LTL formula \(\varphi\) and trace \(\tau\), \(\tau \models \varphi\) iff
        \(V(\varphi, \tau) > 0\).
    \end{corollary}

    \begin{proof}
        \uline{Forward Direction \(\Rightarrow\):}

        Given an LTL formula \(\varphi\) and trace \(\tau\) such that \(\tau
        \models \varphi\), we can construct a satisfying assignment to the
        constraint system \(\Phi^\sample_\depth\) by
        Theorem~\ref{theorem:completeness} where \(\sample = (\{\tau\}, \emptyset)\)
        and \(\depth\) is the depth of \(\varphi\). Further, the theorem
        guarantees \(y_{1, 0}^\tau = V(\varphi, \tau) > 0\). This is the
        required condition.

        \uline{Backward Direction \(\Leftarrow\):}

        Given an LTL formula \(\varphi\) and trace \(\tau\) such that
        \(V(\varphi, \tau) > 0\), construct the constraint system
        \(\Phi^\sample_\depth\) as before. Iteratively assign to each \(y_{i,
        t}^\tau\) the value \(V(\varphi_i, \tau[t:])\), where \(\varphi_i\) is
        the subformula at node \(i\) in the parse tree. After assignment with
        constraints, \(y_{1, 0}^\tau = V(\varphi, \tau) > 0\) as given,
        satisfying the constraint. By Theorem~\ref{theorem:soundness}, \(\tau
        \models \varphi\) as required.


    \end{proof}

\subsection{Optimized Pattern Matching} 
\label{sec:opm}
We now present a variation of Algorithm \ref{algo:opt} where the input consists
of a sample along with  a user-provided formula pattern, where the propositional
variables are unknown. This approach is in the spirit of
\cite{lemieux2015generalTexada}, where a formula template is used instead of just a depth as in Algorithm \ref{algo:opt}. We generate a static syntax tree by
parsing the given pattern, with pattern propositional variables becoming the
leaves. In addition to the constraints discussed in Section \ref{algo:opt},
constraint \eqref{eq:map} ensures the mapping of the pattern 
variables to exactly one variable in the given sample, where
$Var$ represents the set of variables in the given pattern and $m_{x,p}$ stands
for the mapping of pattern variable $x$ to the sample variable $p$.
\vspace{-0.3em}
\begin{align}
  \textstyle
    \varphi^{\var}_\prop {=} \left [\bigwedge\limits _{x \in Var} \bigvee\limits _{p \in \prop} m_{x, p} \right] \!\! \land \!\!
        \left [\bigwedge\limits _{
        \begin{smallmatrix} x \in Var \\ p \not = p' \in \prop \end{smallmatrix}
        } 
           \hspace{-1mm}
            \!\!\!\!\!\!\neg m_{x, p} \!\!\lor\! \neg m_{x, p'} \right] \label{eq:map}
\end{align}
The constraint (\ref{eq:map}) specifies that each pattern variable $x$ is mapped
to a sample variable, and no pattern variable is mapped to two sample variables. For an example of how this approach and the one defined in the previous subsection is combined into a hybrid approach, which is presented as follow.


\begin{algorithm}[t]
    \begin{algorithmic}[1]
        \Procedure{CompRank}{$\sample = (P, N), depth$} \\ 
        \Comment{\textcolor{red!50!black}{Returns list of
            satisfying formulae $\mathcal{F}$ sorted by score}} 
            \State curr\_depth $\leftarrow$
            0, $\mathcal{F}$ $\leftarrow$ \{literals in $\sample$\} \State used
            $\leftarrow$ \{\}, unary $\leftarrow$ $\{\globally, \finally\}$,
            binary $\leftarrow$ $\{\land, \lor\}$
        
        \While{$curr\_depth \not = depth$} 
            \ForAll{f $\in \mathcal{F}$} \If {$\finally(f)$ does not hold on
            any trace} \\ \Comment{\textcolor{red!50!black}{F-Check}} \State $\mathcal{F}$ $\leftarrow$ $\mathcal{F}$
            $\setminus$ \{f\} \EndIf \EndFor \State used ${\leftarrow}$ used
            $\union$ $\mathcal{F}$ \If{curr\_depth $\not=$ depth - 1} \State $\mathcal{F}$
            $\leftarrow$ $\bigcup_{T \in \textnormal{unary}}
            T(\textnormal{used}) \:\union\: \bigcup_{ T \in \textnormal{binary}}
            T(\textnormal{used}, \textnormal{used})$ \EndIf \State curr\_depth
            $\leftarrow$ curr\_depth + 1 \EndWhile \State scores $\leftarrow$
            $\{(f, V(f, P, N)) | f \in \mathcal{F}\}$ \State
            \textbf{return} sort(scores) \Comment{\textcolor{red!50!black}{sort list w.r.t. scores}}
            \EndProcedure
    \end{algorithmic}
    \caption{Compositional Ranking}
    \label{alg:composition}
\end{algorithm}

\subsection{Hybrid Pattern Matching}
\label{sec:hybrid}

\begin{algorithm}[h]
    \begin{algorithmic}[1]
        \Procedure{HybridPattern}{$\sample = (P, N), pattern$}\\
        \Comment{\textcolor{red!50!black}{ Returns optimal
            formula fitting a pattern}}
        
            \State \textbf{parse} pattern 
            \State construct $\varphi^{\var}_\prop$ for propositional patterns in tree
            \State construct $\varphi^{ST}_\textnormal{n}$ for every subformula pattern $\varphi(n)$ in the tree
            \State constraint $\Phi \leftarrow \varphi^{\var}_\prop \land \bigwedge \varphi^{ST}_\textnormal{n}$
            \State maximize
            $\min( \{y_{1,0}^\tau\: |\: \tau \in P\} )$ with $\Phi$ as constraint
            
            \If{optimization succeeds with model m} \\
            \Comment{\textcolor{red!50!black}{satisfiable}} \State
                construct formula tree from m \State \textbf{return}
                optimized formula tree \Else \State \textbf{return} UNSAT \EndIf
                \EndProcedure
    \end{algorithmic}
    \caption{Computing the optimal formula given a partial pattern}
    \label{algo:hybrid}
\end{algorithm}

The algorithms defined in  sections \ref{sec:cso} and \ref{sec:opm} suffer from
a common drawback, though at different ends of the spectrum. In the
first, we work with increasing depth to find the optimal formula and constraint sizes may grow quickly as a result. In the second,
we start with a formula template and many formulae are
not considered since we are guided by the template pattern. This makes this
approach  \emph{insufficiently} expressive in comparison with
constrained system optimization. 

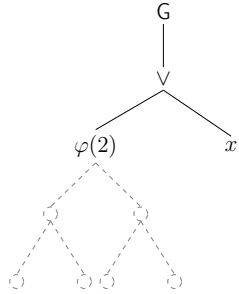
\begin{figure}[t] 
  \centering
\scalebox{.6}{
    \centering
    \begin{tikzpicture}[level/.style={sibling distance=60mm/#1}]
        \node(g){\Large{$\globally$}} child {node (u) {\Large{$\lor$}} child {node (s) {\Large{$\varphi(2)$}} child[dashed, gray]
            {node[draw, circle, minimum size=0.3cm, gray] (l) {} child
            {node[draw, circle, minimum size=0.3cm, gray] (ll) {}} child
            {node[draw, circle, minimum size=0.3cm, gray] (lr) {}}}
            child[dashed, gray] {node[draw, circle, minimum size=0.3cm] (r) {}
            child {node[draw, circle, minimum size=0.3cm, gray] (rl)
            {}} child {node[draw, circle, minimum size=0.3cm, gray]
            (rr) {}}}} child {node
            (x) {\Large{$x$}}}};
      \end{tikzpicture}}
    \caption{Subtree for hybrid pattern matching.}
    \label{fig:subtree}
\end{figure}

To remedy this, we introduce a middle ground, where, instead of attempting to
learn formulae from scratch or from explicit patterns, we learn
\emph{subformulae} within some pattern.
A subformulae argument $\varphi(\depth)$ with $\depth$ being a prescribed
maximum depth for the subtree is provided as part of the pattern, parsed into
the tree as an abstracted empty formula with constraints constructed for the
specified nodes explicitly, and for the subformulae recursively in the manner as
described in Algorithm \ref{algo:hybrid}.  In Figure
\ref{fig:subtree}, we show an example of the hybrid pattern $\globally
(\varphi(2) \: \lor \: x)$, where $\varphi(2)$ is an unknown formula of depth
$\leq 2$ and $x$ is an unknown proposition.

\subsection{Compositional Ranking}
\label{sec:composition}
We describe an alternative greedy search for optimal formula, which
bypasses constraint solving and optimizations, by pruning the search space of
formulae. We begin by enumerating all formulae of depth zero, i.e., all literals
in our system as obtained after parsing input traces. We consider all
compositions of these literals with the operators present. After enumerating the
literals, we  perform an ``$\finally$-check'' :  for any  $\varphi$,
the $\finally$-check tests whether, in any  input sample, $\finally
\varphi$ holds. If a formula passes an $\finally$-check, it is retained to
produce formulae of higher depth, else it is removed  (Algorithm
\ref{alg:composition}). 
We have implemented more custom options to prioritise
certain parts of the search space (Section~\ref{sec:priority-var}).

\subsection{Prioritize Variables. }
    \label{sec:priority-var}
        Finally, we have added one more heuristic in our implementation. 
    In case there is a large set of events and we want to bias the focus of our search
    towards certain letters in the traces that do not occur very often, we may adjust the value $V(p,w)$ assigned to each propositional variable $p$.
    In our default scheme, we assign $V(p,w) = 1$ if $w(1)$ contains $p$.
    A user may assign a value greater than $1$ to variables $p$ that are desirable
    and assign less than $1$ for the variables $p$ that are not.
    This allows for mining specifications pertaining to the prioritized variables in
    cases where several competing well ranked specifications are present.
    Let $\pi$ be the map from the propositional variables $\prop$ to their priority.
    We replace equation~\eqref{eq:prop_score} by the following formula where we
    return score $\pi(p)$ instead of $1$. 
    %
    \begin{gather}
      \textstyle
        \bigwedge _{1 \leq i \leq N} \bigwedge _{p \in \prop}\;\; x_{i, p} \rightarrow \qquad\qquad\nonumber\\
        \qquad\qquad\left[ 
            \bigwedge _{1 \leq t \leq |\tau|} y^\tau _{i, t} = \begin{cases}
                \pi(p) & \textnormal{if } p \in \tau (t) \\
                0 & \textnormal{if } p \not \in \tau (t)
            \end{cases}
        \right] \label{eq:prop_priority_score} 
    \end{gather}

    In the Dining Philosophers problem (see Section~\ref{sec:dining}), we may wish to verify individually
    whether the properties are being satisfied for a single philosopher
    (thread). By giving a higher weight to the properties of this philosopher,
    we can guide the tool to learn the relevant properties and verify them. This
    can be expanded to studying specific applications or threads in varied noisy
    data where the target of interest is either known apriori or is inferred
    from preliminary unguided results. 
    
    
    The suggested variations and their results indicate that our method is
    viable to be adapted to an application at hand, where we want to bias our
    ranking to give preference to a desired class of formulae.

%
%

\section{Experiments}
\label{sec:experiments}
We have implemented the preceding algorithms in a tool called~\ourtool.
In this section, we present the results of \ourtool{} on a set of
traces sampled from a grid-world environment running under OpenAI Gym.
\ourtool{} is implemented in C++. 
For the optimization, it takes a set of positive traces, a (possibly empty) set
of negative traces, and a formula template (which can simply be $\varphi(d)$, a
search depth of $d$ with no specification) or a combination, while the
compositional ranking takes as input the traces along with a maximum search depth.
Our implementation uses SMT solver Z3~\cite{z3} for the optimizations. 
%
All our queries to Z3 are quantifier-free.
For optimization, \ourtool{} returns a formula with maximal score according to our 
scheme, while for compositional ranking, it returns a list of all satisfying formulae in the search space, sorted by score.
In our experiments, we used a discount factor $\alpha = e^{-1}$ and also used $\beta = 0.8$
to decay each time we build deeper formulae in order to bias the ranking towards
simpler formulae.
We evaluated the performance of \ourtool{} on a 64-bit Linux system with an AMD
Renoir Ryzen 5 (4500U) laptop CPU.
We set $1000$ seconds as timeout. We compare with Texada\cite{TexadaTool} and the SAT based tool Traces2LTL \cite{ltlFMCAD18}. 

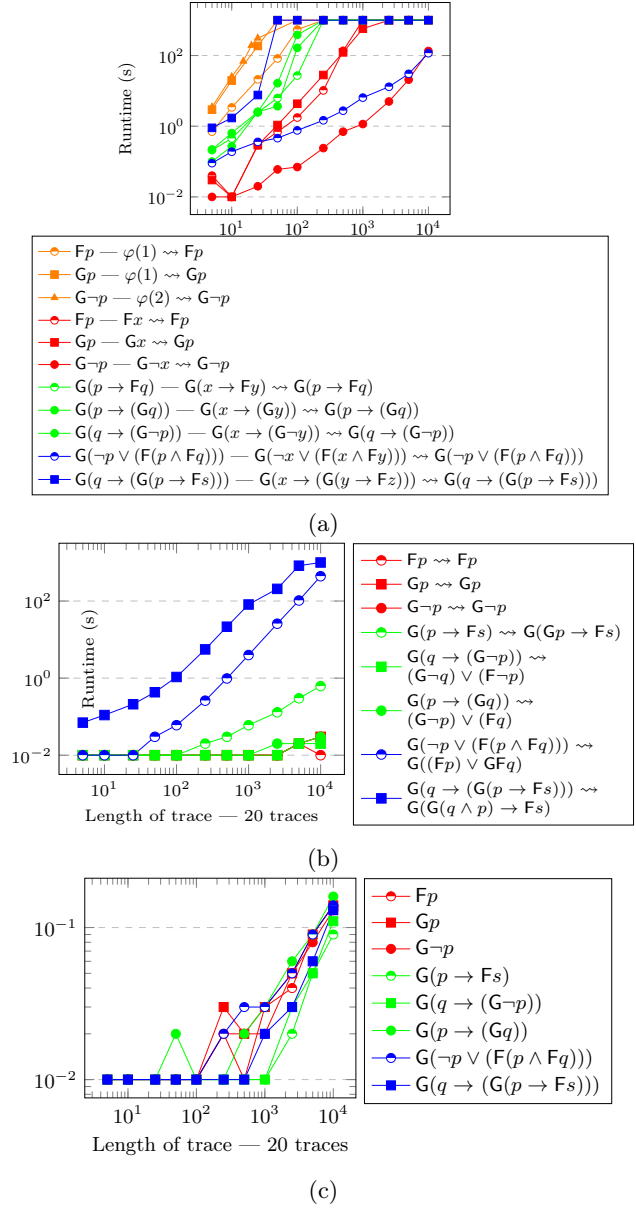
\begin{figure}[hp]
  \centering
  \renewcommand{\thesubfigure}{\alph{subfigure}} 
  \begin{subfigure}[t]{\linewidth}
    \centering
    \scalebox{.8}{\scalebox{0.9}{
\begin{tikzpicture}
        \centering
    
        \begin{axis}[
            xlabel={Length of trace --- 20 traces},
            ylabel={Runtime (s)},
            ymode=log, xmode=log,
            height=5.5cm,
            ymajorgrids=true,
            scaled x ticks = false,
            grid style=dashed,
            legend style={cells={align={left}}, at={(0.5,-0.1)},anchor=north},
            legend cell align={left},
        ]
        \addplot[
                color=orange,
                mark=halfcircle*,
                ]
                coordinates {
                (5, 0.7)
                (10, 3.4)
                (25, 21.2)
                (50, 82.49)
                (100, 542.6)
                (250, 1000)
                (500, 1000)
                (1000, 1000)
                (2500, 1000)
                (5000, 1000)
                (10000, 1000)
                };
        \addlegendentry{$\finally p$ --- $\varphi(1) \rightsquigarrow \finally p$}
        \addplot[
                color=orange,
                mark=square*,
                ]
                coordinates {
                (5, 2.93)
                (10, 19.59)
                (25, 186.6) 
                (50, 1000) 
                (100, 1000)
                (250, 1000)
                (500, 1000)
                (1000, 1000)
                (2500, 1000)
                (5000, 1000)
                (10000, 1000)
                };
        \addlegendentry{$\globally p$ --- $\varphi(1) \rightsquigarrow \globally p$}
        \addplot[
                color=orange,
                mark=triangle*,
                ]
                coordinates {
                (5, 3.49) 
                (10, 25.26)
                (15, 68.78)
                (20, 193)
                (25, 310)
                (100, 1000)
                (250, 1000)
                (500, 1000)
                (1000, 1000)
                (2500, 1000)
                (5000, 1000)
                (10000, 1000)
                };
        \addlegendentry{$\globally \neg p$ --- $\varphi(2) \rightsquigarrow \globally \neg p$}
        \addplot[
                color=red,
                mark=halfcircle*,
                ]
                coordinates {
                (5, 0.04)
                (10, 0.01)
                (25, 0.29)
                (50, 0.69)
                (100, 1.76)
                (250, 10.3)
                (500, 136)
                (1000, 1000)
                (2500, 1000)
                (5000, 1000)
                (10000, 1000)
                };
        \addlegendentry{$\finally p$ --- $\finally x \rightsquigarrow \finally p$}
        \addplot[
                color=red,
                mark=square*,
                ]
                coordinates {
                (5, 0.03)
                (10, 0.01)
                (25, 0.29)
                (50, 1.08)
                (100, 4.32)
                (250, 28.07)
                (500, 123)
                (1000, 577)
                (2500, 995)
                (5000, 1000)
                (10000, 1000)
                };
        \addlegendentry{$\globally p$ --- $\globally x \rightsquigarrow \globally p$}
        \addplot[
                color=red,
                mark=*,
                ]
                coordinates {
                (5, 0.01)
                (10, 0.01)
                (25, 0.02)
                (50, 0.06)
                (100, 0.07)
                (250, 0.24)
                (500, 0.70)
                (1000, 1.15)
                (2500, 5.01)
                (5000, 20.75)
                (10000, 134.16)
                };
        \addlegendentry{$\globally \neg p$ --- $\globally \neg x \rightsquigarrow \globally \neg p$}
        \addplot[
                color=green,
                mark=halfcircle*,
                ]
                coordinates {
                (5, 0.21)
                (10, 0.45)
                (25, 2.45)
                (50, 6.27)
                (100, 27.11)
                (250, 1000)
                (500, 1000)
                (1000, 1000)
                (2500, 1000)
                (5000, 1000)
                (10000, 1000)
                };
        \addlegendentry{$\globally (p \rightarrow \finally q)$ --- $\globally (x \rightarrow \finally y) \rightsquigarrow \globally (p \rightarrow \finally q)$}
        \addplot[
                color=green,
                mark=*,
                ]
                coordinates {
                (5, 0.22)
                (10, 0.64)
                (25, 2.55)
                (50, 16.51)
                (100, 383.11)
                (250, 1000)
                (500, 1000)
                (1000, 1000)
                (2500, 1000)
                (5000, 1000)
                (10000, 1000)
                };
        \addlegendentry{$\globally (p \rightarrow (\globally q))$ --- $\globally (x \rightarrow (\globally y)) \rightsquigarrow \globally (p \rightarrow (\globally q))$}
        \addplot[
                color=green,
                mark=*,
                ]
                coordinates {
                (5, 0.10)
                (10, 0.27)
                (25, 2.55)
                (50, 3.67)
                (100, 164.83)
                (250, 1000)
                (500, 1000)
                (1000, 1000)
                (2500, 1000)
                (5000, 1000)
                (10000, 1000)
                };
        \addlegendentry{$\globally (q \rightarrow (\globally \neg p))$ --- $\globally (x \rightarrow (\globally \neg y))\rightsquigarrow \globally (q \rightarrow (\globally \neg p))$}
        \addplot[
                color=blue,
                mark=halfcircle*,
                ]
                coordinates {
                (5, 0.09)
                (10, 0.19)
                (25, 0.36)
                (50, 0.46)
                (100, 0.76)
                (250, 1.47)
                (500, 2.75)
                (1000, 6.44)
                (2500, 13.06)
                (5000, 30.32)
                (10000, 116.57)
                };
        \addlegendentry{$\globally (\neg p \lor (\finally (p \land \finally q)))$ --- $\globally (\neg x \lor (\finally (x \land \finally y))) \rightsquigarrow \globally (\neg p \lor (\finally (p \land \finally q)))$}
        \addplot[
                color=blue,
                mark=square*,
                ]
                coordinates {
                (5, 0.90)
                (10, 1.71)
                (25, 7.63)
                (50, 1000)
                (100, 1000)
                (250, 1000)
                (500, 1000)
                (1000, 1000)
                (2500, 1000)
                (5000, 1000)
                (10000, 1000)
                };
        \addlegendentry{$\globally (q \rightarrow (\globally (p \rightarrow \finally s)))$ --- $\globally (x \rightarrow (\globally (y \rightarrow \finally z))) \rightsquigarrow \globally (q \rightarrow (\globally (p \rightarrow \finally s)))$}
    \end{axis}
    \end{tikzpicture}
   } 
    
    }
    \caption{}
    \label{subfig:runtime_opt}
  \end{subfigure}
  \begin{subfigure}[t]{\linewidth}
    \centering
    \scalebox{1}{\scalebox{0.9}{
\begin{tikzpicture}
    \centering
    \scriptsize
    \begin{axis}[
        xlabel={Length of trace --- 20 traces},
        ylabel={Runtime (s)},
        y label style={at={(axis description cs:0.16,.5)},anchor=south},
        ymode=log, xmode=log,
        ymajorgrids=true,
        scaled x ticks = false,
        grid style=dashed,
        legend pos=outer north east,
        legend style={cells={align={left}}},
        legend cell align={left},
        height=5cm,
    ]
    \addplot[
            color=red,
            mark=halfcircle*,
            ]
            coordinates {
            (5, 0.01)
            (10, 0.01)
            (25, 0.01)
            (50, 0.01)
            (100, 0.01)
            (250, 0.01)
            (500, 0.01)
            (1000, 0.01)
            (2500, 0.01)
            (5000, 0.02)
            (10000, 0.01)
            };
    \addlegendentry{$\finally p \rightsquigarrow \finally p$}
    \addplot[
            color=red,
            mark=square*,
            ]
            coordinates {
            (5, 0.01)
            (10, 0.01)
            (25, 0.01)
            (50, 0.01)
            (100, 0.01)
            (250, 0.01)
            (500, 0.01)
            (1000, 0.01)
            (2500, 0.01)
            (5000, 0.02)
            (10000, 0.03)
            };
    \addlegendentry{$\globally p \rightsquigarrow \globally p$}
    \addplot[
            color=red,
            mark=*,
            ]
            coordinates {
            (5, 0.01)
            (10, 0.01)
            (25, 0.01)
            (50, 0.01)
            (100, 0.01)
            (250, 0.01)
            (500, 0.01)
            (1000, 0.01)
            (2500, 0.01)
            (5000, 0.02)
            (10000, 0.03)
            };
    \addlegendentry{$\globally \neg p \rightsquigarrow \globally \neg p$}
    \addplot[
            color=green,
            mark=halfcircle*,
            ]
            coordinates {
            (5, 0.01)
            (10, 0.01)
            (25, 0.01)
            (50, 0.01)
            (100, 0.01)
            (250, 0.02)
            (500, 0.03)
            (1000, 0.06)
            (2500, 0.13)
            (5000, 0.30)
            (10000, 0.62)
            };
    \addlegendentry{$\globally (p \rightarrow \finally s) \rightsquigarrow \globally (\globally p \rightarrow \finally s)$}
    \addplot[
            color=green,
            mark=square*,
            ]
            coordinates {
            (5, 0.01)
            (10, 0.01)
            (25, 0.01)
            (50, 0.01)
            (100, 0.01)
            (250, 0.01)
            (500, 0.01)
            (1000, 0.01)
            (2500, 0.01)
            (5000, 0.02)
            (10000, 0.02)
            };
    \addlegendentry{$\globally (q \rightarrow (\globally \neg p)) \rightsquigarrow $ \\ $ (\globally \neg q) \lor (\finally \neg p)$}
    \addplot[
            color=green,
            mark=*,
            ]
            coordinates {
            (5, 0.01)
            (10, 0.01)
            (25, 0.01)
            (50, 0.01)
            (100, 0.01)
            (250, 0.01)
            (500, 0.01)
            (1000, 0.01)
            (2500, 0.02)
            (5000, 0.02)
            (10000, 0.03)
            };
    \addlegendentry{$\globally (p \rightarrow (\globally q)) \rightsquigarrow $ \\ $ (\globally \neg p) \lor (\finally q)$}
    \addplot[
            color=blue,
            mark=halfcircle*,
            ]
            coordinates {
            (5, 0.01)
            (10, 0.01)
            (25, 0.01)
            (50, 0.03)
            (100, 0.06)
            (250, 0.26)
            (500, 0.98)
            (1000, 3.97)
            (2500, 25.81)
            (5000, 103.27)
            (10000, 440.32)
            };
    \addlegendentry{$\globally (\neg p \lor (\finally (p \land \finally q))) \rightsquigarrow $ \\ $\globally( (\finally p) \lor \globally \finally q)$}
    \addplot[
            color=blue,
            mark=square*,
            ]
            coordinates {
            (5, 0.07)
            (10, 0.11)
            (25, 0.21)
            (50, 0.43)
            (100, 1.07)
            (250, 5.56)
            (500, 21.55)
            (1000, 81.31)
            (2500, 207.00)
            (5000, 825.76)
            (10000, 1000)
            };
    \addlegendentry{$\globally (q \rightarrow (\globally (p \rightarrow \finally s))) \rightsquigarrow $ \\ $\globally (\globally(q \land p) \rightarrow \finally s)$}
\end{axis}
\end{tikzpicture}

}
} 
    \caption{}    
    \label{subfig:runtime_comp}
  \end{subfigure}\\
  \begin{subfigure}[t]{\linewidth}
    \centering
    \scalebox{.95}{\scalebox{0.9}{
\begin{tikzpicture}
    \centering

    \begin{axis}[
        xlabel={Length of trace --- 20 traces},
        height=5cm,
        ymode=log, xmode=log,
        ymajorgrids=true,
        scaled x ticks = false,
        grid style=dashed,
        legend pos=outer north east,
        legend cell align={left},
    ]
    \addplot[
            color=red,
            mark=halfcircle*,
            ]
            coordinates {
            (5, 0.01)
            (10, 0.01)
            (25, 0.01)
            (50, 0.01)
            (100, 0.01)
            (250, 0.02)
            (500, 0.01)
            (1000, 0.03)
            (2500, 0.04)
            (5000, 0.09)
            (10000, 0.14)
            };
    \addlegendentry{$\finally p$}
    \addplot[
            color=red,
            mark=square*,
            ]
            coordinates {
            (5, 0.01)
            (10, 0.01)
            (25, 0.01)
            (50, 0.01)
            (100, 0.01)
            (250, 0.03)
            (500, 0.02)
            (1000, 0.03)
            (2500, 0.05)
            (5000, 0.09)
            (10000, 0.14)
            };
    \addlegendentry{$\globally p$}
    \addplot[
            color=red,
            mark=*,
            ]
            coordinates {
            (5, 0.01)
            (10, 0.01)
            (25, 0.01)
            (50, 0.01)
            (100, 0.01)
            (250, 0.02)
            (500, 0.02)
            (1000, 0.02)
            (2500, 0.05)
            (5000, 0.08)
            (10000, 0.13)
            };
    \addlegendentry{$\globally \neg p$}
    \addplot[
            color=green,
            mark=halfcircle*,
            ]
            coordinates {
            (5, 0.01)
            (10, 0.01)
            (25, 0.01)
            (50, 0.01)
            (100, 0.01)
            (250, 0.01)
            (500, 0.01)
            (1000, 0.01)
            (2500, 0.02)
            (5000, 0.05)
            (10000, 0.09)
            };
    \addlegendentry{$\globally (p \rightarrow \finally s)$}
    \addplot[
            color=green,
            mark=square*,
            ]
            coordinates {
            (5, 0.01)
            (10, 0.01)
            (25, 0.01)
            (50, 0.01)
            (100, 0.01)
            (250, 0.01)
            (500, 0.01)
            (1000, 0.01)
            (2500, 0.03)
            (5000, 0.05)
            (10000, 0.11)
            };
    \addlegendentry{$\globally (q \rightarrow (\globally \neg p))$}
    \addplot[
            color=green,
            mark=*,
            ]
            coordinates {
            (5, 0.01)
            (10, 0.01)
            (25, 0.01)
            (50, 0.02)
            (100, 0.01)
            (250, 0.01)
            (500, 0.02)
            (1000, 0.03)
            (2500, 0.06)
            (5000, 0.09)
            (10000, 0.16)
            };
    \addlegendentry{$\globally (p \rightarrow (\globally q))$}
    \addplot[
            color=blue,
            mark=halfcircle*,
            ]
            coordinates {
            (5, 0.01)
            (10, 0.01)
            (25, 0.01)
            (50, 0.01)
            (100, 0.01)
            (250, 0.02)
            (500, 0.03)
            (1000, 0.03)
            (2500, 0.05)
            (5000, 0.09)
            (10000, 0.14)
            };
    \addlegendentry{$\globally (\neg p \lor (\finally (p \land \finally q)))$}
    \addplot[
            color=blue,
            mark=square*,
            ]
            coordinates {
            (5, 0.01)
            (10, 0.01)
            (25, 0.01)
            (50, 0.01)
            (100, 0.01)
            (250, 0.01)
            (500, 0.01)
            (1000, 0.02)
            (2500, 0.03)
            (5000, 0.06)
            (10000, 0.13)
            };
    \addlegendentry{$\globally (q \rightarrow $  $(\globally (p \rightarrow \finally s)))$}
\end{axis}
\end{tikzpicture}
}

}
    \caption{}
    \label{subfig:runtime_tex}
  \end{subfigure}\vspace{-1em}
  \caption{Runtime for traces generated for formulae. (a)full and partial
    pattern specification. (b)  compositional ranking, and only a depth as
    input. (c)  Texada~\cite{TexadaTool}, with complete pattern specification.
    In the legend of (a), $ \varphi$ --- $pattern \rightsquigarrow \psi$
    represents the following : $\varphi$ is the formula used to generate the
    traces; $pattern$ is the input to \ourtool{} and $\psi$ is the learned
    output formula. Similarly in the legend of
    (b), $\varphi \rightsquigarrow \psi$ indicates the traces in the sample were
    generated using $\varphi$, and $\psi$ is the learnt formula.  
  }
  \label{fig:runtime}
\end{figure}

    Due to the vastly different mechanisms of the tools and algorithms in contention, an apples-to-apples comparison is impossible. The compositional ranking input accepts no pattern input, constraint system optimization accepts none to partial input, while Texada requires complete template specification as input, and does not utilize negative trace samples either. We defer the comparison with Traces2LTL to a later point in the section, as it accepts a different method of trace input, i.e., rational traces. As a whole, \ourtool{} is capable of reliably extracting properties from limited, noisy data, and with the addition of compositional ranking, is able to match the performance of existing tools while requiring no guidance for inference. The results are summarized in Figure~\ref{fig:runtime}. 

    When comparing to Traces2LTL, since both it and our constraint system optimization encode a similar system, the former restricted to just boolean variables instead of real scores, for large trace inputs as tested above, both output similar results, and due to avoiding searching real parameters, Traces2LTL outperforms \ourtool{}. 
    For smaller noisy traces, with effective number of trace events ranging approximately from 20 to 500, Traces2LTL outputs several possible formulae, most containing the extraneous noisy variables, since it relies on Z3 to present the formulae in arbitrary orders as satisfying assignments. In this scenario, \ourtool{} reliably bypasses noise and extracts formulae of up to depth 3 reliably, and we suggest this as the ideal use-case for constraint system optimization over other algorithms. 
    
    Traces2LTL \cite{ltlFMCAD18} also presents an alternate method of using decision-tree learning to remedy scalability issues of SMT solver based systems for larger inputs. However, compositional ranking is several orders of magnitude faster than either of the two methods for higher depth formulae while producing more succinct specifications, and is thus suggested as an alternative in that case.
    
    Due to the different natures of inputs for the programs, we split the tests to compare against each. We compare performance with the simpler Texada and reliability against the expressive Traces2LTL.
    
    For the NMRDP tests with Traces2LTL, given the input, it was asked to iteratively produce 5 satisfying assignments, and the ILE was averaged over these results, since the tool does not have a preferential ordering scheme. In several cases, the correct generator was one of the 5 assignments presented, which scored an ILE of 0, while the others, depending on compatibility with the sample and closeness to original, were spread between an ILE of 0.3 and 0.8 for randomized grids. 
    
    For \ourtool{}, with both algorithms, one of the formulas $(\globally (p\Rightarrow \finally q))$ was learnt as a close alternative (different for both algorithms too!), leading to an increased ILE of 0.25-0.35 in that case, while the other formulas were learnt exactly. The results were averaged over runs and formulas for all three.
    
    Traces2LTL was considered for these tests over Texada to maintain a level playing field regarding guidance. Since for Texada, to obtain the correct formula, one must know and provide the entire template.
    
    Due to scalability issues of the SAT/SMT based methods, both Traces2LTL and \ourtool{} with constraint optimization were provided a randomly chosen subset of the generated traces, while \ourtool{} with compositional ranking processed the entire set.

%





\begin{figure}[h]
    \centering
    \begin{tabular}{c|c|c|c}
         & Constraint & Comp. & Traces2LTL\\
         & Optimization & Ranking & \\
         \hline 
         Mean ILE & 0.031 & 0.037 & 0.112\\
         Input size & $10^3$ & $2\times 10^5$ &$10^3$
    \end{tabular}
    \label{fig:my_label}
\end{figure}

We consider two synthetic tests. First, we systematically generate  trace of different sizes to compare performance and expressiveness of the different methods. After generating uniformly random traces, we add noise to each point in the trace in the form of two extraneous propositional variables with probability \(p_{\text{noise}}\) each. For these experiments, we maintained \( p_{\text{noise}} = 0.25\).  We consider LTL formulae encoding some popular requirements. 

\subsection{Non-Markovian IRL} We apply learning techniques to generate a reward function over an MDP defined as a grid world to obtain an NMRDP, similar to the example in Figure~\ref{fig:syntax}(a). After generating a randomized $10\times 10$ grid environment labeled with propositional variables, we uniformly sample the grid taking actions compatible with an input automaton. This ensures that the generated traces satisfy a given formula. We use the same LTL properties as the previous case. We allow the MDP to randomly simulate for at least 100 steps, after which we wait for it to reach an accepting state. Through this method, we generated traces of length varying between 100 and 150, with 1000 positive and 1000 negative traces for each formula,  amounting to  a total trace length of at least $10^5$ across all positive and negative inputs. However, for constraint system optimization and Traces2LTL \cite{ltlFMCAD18}, due to timeouts, a smaller subset was randomly selected from the traces.

For our experiments, given an NMRDP $M$ in the form of a gridworld, we can compute the optimal policy using the preceding equation for three different DRA(Definition \ref{DRA} in Section \ref{sec:product-construction}) objectives by computing three product MDPs. The first DRA objective  is what we are trying to learn. We will denote the optimal policy here as $\pi^*_\text{true}$ computed on $M \times A_\text{true}$, where $A_\text{true}$ is the DRA representation of the LTL objective we are trying to learn. Then, we have the policy $\pi^*_\text{QL}$ computed on $M \times A_\text{QL}$, where $A_\text{QL}$ is the DRA learnt using \ourtool{}. Finally, we have the policy $\pi^*_\text{T2L}$ computed on $M \times A_\text{T2L}$, where $A_\text{T2L}$ is the DRA learnt using Traces2LTL \cite{ltlFMCAD18}. We will take these three policies to generate our results in the form of the inverse learning error (ILE). We compute these value functions using uniformly random sampling of trajectories from every state in the NMRDP. We can then take a simple ratio (MeanILE in Table below) of the number of trajectories  satisfied by $A_\text{true}$, and  divide it with the total number of trajectories, and report an average over multiple runs and inputs. In particular, we will compute two ILE values, comparing 
$|| V_{\pi^*_\text{true}} - V_{\pi^*_\text{QL}} ||_2$ and $|| V_{\pi^*_\text{true}} - V_{\pi^*_\text{T2L}} ||_2$. Our experiments demonstrate that the former is smaller than the latter, thereby providing evidence that our approach generalizes better for non-Markovian IRL than a competing one adapted to the IRL.


\subsection{Runtimes with optimization for scores} We report the average runtimes for our approach with and without compositional ranking and compare it to that of Texada. In Figure~\ref{subfig:runtime_opt}, we show the runtimes with different patterns of user specification over the different lengths of traces for constraint system optimization and optimized pattern matching. 
In Figure~\ref{subfig:runtime_comp} we show runtimes for synthetic traces generated from the following common formulae 
with compositional ranking~\cite{ltlFMCAD18,dwyer1998property}.

\begin{center}
    \begin{tabular}{ c | c }
        Property    & Formulae \\
        \hline
        Absence     & $\globally \neg p$,
                     $\globally (q \rightarrow \globally (\neg p))$ \\

        Response    & $\globally (p \rightarrow \finally s)$,
                     $\globally (q \rightarrow \globally (p \rightarrow
                    \finally s))$ \\

        Existence   &  $\finally p$, 
                     $\globally (\neg p \lor \finally (p \land \finally q))$ \\

        Universality& $\globally p$, 
                     $\globally (p \rightarrow \globally q)$ \\
    \end{tabular}
    \end{center}

\vspace{0.5em}The length of the individual traces  varies from 5 to 10,000. In Figure~\ref{subfig:runtime_tex} we show runtimes for the generated traces with Texada. 
The comparison with Texada is difficult because it requires the complete formula template as input while we do not. 
The tests are thus performed with complete formula template as input, with which Texada outputs a list of possible formulae with propositions substituted in. 
In our eperiments, we found that Texada and compositional ranking perform comparably despite the fact that our approach requires no pattern input.

\subsection{Non-Markovian Rewards and Product Construction}
\label{sec:product-construction}
Since the non-Markovian nature of the reward signal in our setting can be represented by a regular or $\omega$-regular language, we translate the problem of finding an optimal policy to one of finding a Markovian optimal policy over an augmented decision process by taking into account the structure of the regular language. Since we are learning LTL formulae, the corresponding structures are deterministic Rabin automata, though other choices of automata are possible.

\begin{definition}[Deterministic Rabin Automaton (DRA)]
    A DRA $\A$ is a tuple $(\Sigma, Q, q_0, \delta, F)$, where $\Sigma$ is a finite \emph{alphabet}, $Q$ is a finite set of \emph{states}, $q_0 \in Q$ is the \emph{initial state}, $\delta : Q \times \Sigma \to Q$ is the transition function, and $F = \{(B_i, G_i) \in 2^Q{\times} 2^Q\}$ is the Rabin condition.
	\label{DRA}
\end{definition}

By taking the DRA representation of a regular reward signal and the underlying NMRDP, a product Markov Decision Process (MDP) can be computed. In particular, the reward signal of the MDP is defined only over the current state and action, thereby enabling the adoption of conventional Markovian solutions, such as value iteration. In order to define a reward signal for the product MDP, we must ensure that, when the acceptance condition of the underlying DRA is satisfied, the agent is rewarded. For this, we must compute maximal sub-MDPs known as maximal end components (MECs). Formally, for a DRA with acceptance condition $F = (G, B)$, a MEC $E = (S^E, A^E)$ of the product MDP $\Mm \times \mathcal{A}$ is accepting if  $S^E \cap (S \times B) = \emptyset$ and $S^E \cap (S \times G) \neq \emptyset$ for some $(B, G) \in F$. The reward signal of the product MDP is defined whenever states with labels in $G$ are entered.

\begin{definition}[Product LMDP]
Given NMRDP $\Mm = (S, s_0, A, T, R, \Sigma, L)$, DRA $\mathcal{A} = (Q, q_0, \Sigma, \delta, F)$, the \emph{product} LMDP $M {\times} A$ is the tuple $(S^\times, (s_0, q_0), A, T^\times, R^\times, L, \Sigma)$, where $S^\times = S \times Q$; $T^\times(s', q' | s, q, a)$ equals $T(s' | s, a)$ if $q' = \delta(q, L(s'))$ and is $0$ otherwise; and $R^\times(s, q, a, s', q') = 1$ if $q' \in G_i$ for some $i$ and is $0$ otherwise. 
\end{definition}

For a product LMDP $(S^\times, (s_0, q_0), A, T^\times, R^\times, L^\times, \Sigma)$, Equation (\ref{equation:expectedReward}) reduces to the equation below, where $r = R(s, q, \pi(s, q), s', q')$.
\begin{equation}
    \nonumber
    V_\pi (s, q) = \sum_{(s', q') \in S^\times} T(s', q' | s, q, \pi(s, q)) \left[ r + \gamma V_\pi (s', q') \right]
    \label{equation:productValue}
\end{equation}
Since we are now reasoning about a Markovian reward over the product, the optimal value for a state $V^* (s, q) = \max_\pi V_\pi (s, q)$ can be computed using value iteration. In particular, we can initialize $V(s, q) = 0$ for all $q \notin G_i$ for some $G_i \in F$ and $V(s, q) = 1$ for all $q \in F$. We can then iteratively refine these value functions by applying the following equation until convergence, where $r = R(s, q, a, s', q')$.
\begin{equation}
    \nonumber
    V (s, q) = \max_a \sum_{(s', q') \in S^\times} T(s', q' | s, q, a) \left[ r + \gamma V (s', q') \right]
    \label{equation:valueIteration}
\end{equation}
These equations converge to $V^* (s, q)$ for all $(s, q) \in S^\times$. The optimal policy can then be extracted as
\begin{equation}
    \nonumber
    \pi^* (s, q) = \arg \! \max_a \!\!\!\!\! \sum_{(s', q') \in S^\times} \!\!\!\!\! T(s', q' | s, q, a) \left[ r + \gamma V^* (s', q') \right]
    \label{equation:optimalPolicy}
\end{equation}






\subsection{Mining Formulae from the traces of Dining Philosophers}
\label{sec:dining}
The dining philosophers problem~\cite{dijkstra1971hierarchical} is a widely used
example of a control problem in distributed systems and has become an important
benchmark for testing expressiveness of concurrent languages and resource
allocation strategies. We consider the problem with  five philosophers
$\texttt{p1}$, $\texttt{p2}$, $\texttt{p3}$, $\texttt{p4}$, $\texttt{p5}$
sitting at a round table. They are being served food, with a fork placed between
each pair. Each philosopher proceeds to think till they are hungry, after which
they attempt to pick up the forks on both of their sides, eating till they are
full, but only when forks on both sides are available. After they are done
eating, they put the forks down back on to either of their sides, and continue
thinking.
The goal is to
establish \emph{lockout-freedom}, i.e., each hungry philosopher is eventually
able to eat.
We use $\ourtool$ to mine LTL formulae from a trace of size 250 from Texada
tests \cite{TexadaTool}.
We searched several mined properties using different templates presented in Figure~\ref{fig:diningphilo}.
%
%
  \begin{itemize}
      \item When we gave the  pattern $\globally(\varphi(1))$ (invariant, depth 1) to
      \ourtool{} and required it to learn a property, we obtained the property
      $\globally \finally \texttt{p1 is thinking}$. The mined property acts as a
      verification of the liveness of the system. \ourtool{} took 12 seconds to find
      the property. While this provides a general fact about the system, building
      upon this result, we can guide the tool to find more relevant properties for
      higher depths. 
      
      \item When we gave the pattern $\globally \: (x \rightarrow \varphi(1))$ to
      \ourtool{} and required it to learn a property, we obtained the property
      $\globally \: ((\texttt{p4 is eating}) \rightarrow \neg (\texttt{p3 is
      eating}))$. The mined property illustrates that adjacent philosophers cannot
      acquire forks at the same time, ensuring that our \emph{lock}, the
      availability of forks does indeed prevent philosophers from eating. \ourtool{}
      took 72 seconds to find the property.
      
      \item  When we gave the pattern $\globally \:(x \rightarrow \finally\: y \land
      \finally\: z)$ to \ourtool{} and required it to learn a property, we obtained
      $\globally \: ((\texttt{p1 is hungry}) \rightarrow (\finally\: ((\texttt{p1 is
      eating})\land \finally\: (\texttt{p1 is thinking}))) ) $. This property is a
      richer demonstration of deadlock freedom for philosopher 1, ensuring that they
      both enter and exit their \emph{critical section}, i.e., the $\texttt{eating}$
      state. \ourtool{} took 165 seconds to find the property.
  \end{itemize}

by observing their runs.
Using patterns, \ourtool{} can be guided to learn properties of specific
interest.

\begin{figure}[t]
  \centering
  \tiny
  \begin{tabular}{| c | c | c |}
   \hline
   Input                                   & Output                                                                            & Interpretation \\
   \hline
   $\globally(\varphi(1))$                    & $\globally \finally \texttt{p1 is thinking}$                                      & Liveness property \\
   \hline
   $\globally \: (x \rightarrow \varphi(1))$  & $\globally \: ((\texttt{p4 is eating}) \rightarrow \neg (\texttt{p3 is eating}))$ & Mutual exclusion \\
   \hline
   $\globally \:(x \rightarrow \finally\: y \land \finally\: z)$ & $\globally \: ((\texttt{p1 is hungry}) \rightarrow$         & Deadlock freedom \\ 
                                           & $(\finally\: ((\texttt{p1 is eating})\land \finally\: (\texttt{p1 is thinking}))) ) $ & \\ 
   \hline
  \end{tabular}
  \vspace{-2mm}
  \caption{Results of running \ourtool{} on Dining Philosophers traces}
  \label{fig:diningphilo}
  \vspace{-8mm}
\end{figure}

\section{Related Work}
\label{sec:related}
\cite{arora2021survey} present a detailed landscape of the algorithms, challenges, and the state-of-the-art in IRL.
\cite{ng2000algorithms} formalized the first computation solution to IRL based on linear programming to demonstrate the effectiveness of IRL.  
Among more recent efforts, IRL is solved using techniques from entropy optimization\cite{ziebart2008maximum,boularias2011relative,haarnoja2018soft},  maximum likelihood estimation~\cite{vroman2014maximum,scobee2019maximum}, and reformulating the problem as a classification task~\cite{klein2012inverse}. 
These techniques have shown much promise and have been applied successfully to problems such as maneuvering remote-controlled helicopters \cite{abbeel2007application} and Atari games \cite{tucker2018inverse}.

Grammatical inference is a related area~\cite{lopes2009active} concerned with learning grammars and their automata representations \cite{grammaticalInference,grammaticalInferenceBook}. Active techniques rely on querying the system under learning to guide the inference process, whereas passive grammatical inference leverages a static set of trace behavior without making further queries for additional data. The former is exemplified by $L^\star$ algorithm \cite{angluin1987learning}, whereas the latter generally relies on state-merging procedures and can be used to learn probabilistic automata \cite{ALERGIA}, MDPs in the context of model checking \cite{IO_ALERGIA_1}, timed automata \cite{IO_ALERGIA_2}, and regular decision processes \cite{AbadiBrafman20}. There is a growing literature on the application of grammatical inference to RL. This typically entails the learning of weighted DFAs, known as reward machines \cite{toro2019learning,xu2020SAT,XuWuNeiderTopcu21}.

However, the application of grammatical inference for IRL has not been explored. Indeed, the foregoing RL methods learn automata from traces of the underlying decision process with a given reward signal, not from traces of an expert policy over unknown and unobservable environment. The problem of learning LTL formulae from traces is a form of grammatical inference that has been well-studied. Two of the methods most related to our own are presented in \cite{lemieux2015generalTexada} and \cite{ltlFMCAD18}. The focus of \cite{ltlFMCAD18} is to produce the minimal formula which is consistent with a rational sample represented as a lasso. The problem of matching formulae with traces is encoded as a constraint system and a satisfying assignment yields the learned property. However, requiring the inputs to be lassos significantly restricts the application to real scenarios. The method in \cite{lemieux2015generalTexada} requires a user-defined input template of the LTL formula which they would like to satisfy and outputs all possible propositional substitutions consistent with the sample.

We seek to combine some of these ideas and try to eliminate their restrictions with a method that may be used on finite traces obtained from real systems, and output results relevant to the user. To allow the algorithms to quantitatively distinguish formulae, we supplement them with the idea of a ranking scheme as a parameter to the methods. Our ranking scheme quantitatively scores each formula against a finite word. 
It expands on intuitive ideas used to formulate distances in regular language spaces \cite{fulop2015topology,parker2016regular}. The suggested scheme assigns a formula a high score if it expresses most features of the word. A formula can score well on a word if it provides longer evidence of validity with respect to larger parts of the word. For example, $\globally a$ will score more on the word $aaaa$ than on the word $aa$, since it contains more evidence for the word to have been a prefix of \(a^\omega\in L(\globally a)\). Furthermore, our suggested scoring scheme encourages simpler formulae over complex ones in equivalence classes; that is, from an equivalence class of LTL formulae, the ones with smaller parse trees will be preferred.

The papers closest to our work are \cite{lemieux2015generalTexada} and \cite{ltlFMCAD18}. The focus of \cite{ltlFMCAD18} is to produce the minimal formula which is consistent with a rational sample irrespective of the expressiveness, while \cite{lemieux2015generalTexada} requires a user defined input template of the LTL formula which they would like to satisfy. Both work with infinite traces: \cite{ltlFMCAD18} takes rational traces in the form $uv^{\omega}$ where $u,v$ are typically words of length $\sim 10$, while \cite{lemieux2015generalTexada} mines specifications from finite traces, and appends them with an infinite sequence of terminal events. \cite{DBLP:conf/otm/AalstBD05t42} considers finite traces, and develops a LTL checker which takes an event log and a LTL property and verifies if the observed behaviour matches some bad behaviour. The papers \cite{989799t19,10.1007/3-540-46002-0_24t20} look at the application of monitoring the execution of Java programs, and check LTL formulae on finite traces of these programs. In \cite{lo2012mining33}, the authors focus on mining quantified temporal rules which help in establishing data flow analysis between variables in a program, while in \cite{weimer2005mining46} software bugs are exposed using a mining algorithm, especially for control flow paths. \cite{agrawal1998miningt9} looks at process mining in the context of workflow management. The tool PISA \cite{zur2000workflowt27} is developed to extract the performance matrix from workflow logs, while a declarative language  is developed to formulate workflow-log properties in \cite{6951474t5}. Tool Synoptic \cite{beschastnikh2011leveraging7} on the other hand, follows a different approach and takes event logs and regular expressions as input and produces a model that satisfies a temporal invariant which has been mined from the trace.


\section{Conclusion and Future work}
\label{sec:conclusion}


In this paper, we presented a novel scheme to quantitatively evaluate  LTL formulae.  Our evaluation schema is designed such that the score received by a word is proportional to how well it represents the formula. Thus, 
words which are ``good representatives''  score higher than words 
which merely satisfy the formula (and hence qualify to be ``poor representatives''). 

One of our contributions is to  use this schema to mine LTL formulae from the traces of reactive systems. Our approach presents a viable solution to non-Markovian inverse reinforcement learning (IRL) in settings where  the reward signal can be captured as  LTL formulae. A possible direction for future work is to 
enhance our optimizer so that it works seamlessly with 
a broad spectrum of constraints.

\bibliographystyle{unsrt} 
\bibliography{biblio}

\end{document}